\documentclass[%
 reprint,
 amsmath,amssymb,
 aps,longbibliography,nobalancelastpage,
]{revtex4-1}

\usepackage{graphicx}
\usepackage{dcolumn}
\usepackage{bm}
\usepackage{adjustbox}
\usepackage{textcomp}
\usepackage{chngcntr}
\usepackage{amsmath,amssymb,amsfonts}
\usepackage{graphicx}
\usepackage[percent]{overpic}
\usepackage{lipsum}
\usepackage{enumerate}
\usepackage{mathrsfs}
\usepackage{balance}
\usepackage{mathtools}
\usepackage{xr}
\usepackage{subfigure}
\usepackage{afterpage}
\usepackage{cases}
\usepackage{scalerel}

\newcommand{\pvec}[1]{\vec{#1}\mkern2mu\vphantom{#1}}
\newcommand\mydots{\ifmmode\ldots\else\makebox[1em][c]{.\hfil.\hfil.}\fi}

\nobalance

\begin{document}

\preprint{APS/123-QED}

\title{A Discrete Fourier Transform-Based Framework for Analysis and Synthesis of Cylindrical Omega-bianisotropic Metasurfaces}

\author{Gengyu Xu}
 \email{paul.xu@mail.utoronto.ca}
\author{George V. Eleftheriades}%
\author{Sean V. Hum}%
\affiliation{%
The Edward S. Rogers Sr. Department of Electrical and Computer Engineering, University of Toronto,\\ Toronto, Ontario M5S 3H7, Canada}%

\date{\today}

\begin{abstract}
This paper presents a framework for analyzing and designing cylindrical omega-bianisotropic metasurfaces, inspired by mode matching and digital signal processing techniques. Using the discrete Fourier transform, we decompose the the electromagnetic field distributions into orthogonal cylindrical modes and convert the azimuthally varying metasurface constituent parameters into their respective spectra. Then, by invoking appropriate boundary conditions, we set up systems of algebraic equations which can be rearranged to either predict the scattered fields of prespecified metasurfaces, or to synthesize metasurfaces which support arbitrarily stipulated field transformations. The proposed framework facilitates the efficient evaluation of field distributions that satisfy local power conservation, which is one of the key difficulties involved with the design of passive and lossless scalar metasurfaces. It represents a promising solution to circumvent the need for active components, controlled power dissipation, or tensorial surface polarizabilities in many state-of-the art conformal metasurface-based devices. To demonstrate the robustness and the versatility of the proposed technique, we design several devices intended for different applications and numerically verify them using finite element simulations.
\end{abstract}

\maketitle

\section{Introduction}
Electromagnetic metasurfaces are devices of subwavelength thickness consisting of two-dimensional arrangements of scatterers (meta-atoms) with engineered electric and/or magnetic polarizabilities and subwavelength dimensions~\cite{Quevedo_Teruel_2019,HMS_review}. Due to their extraordinary ability to efficiently manipulate various aspects of electromagnetic waves, metasurfaces have been leveraged in a variety of practical applications. For example, beam redirectors and splitters~\cite{HMS1,beam_redirection1,DBMG,auxiliary,Chen_refraction}, flat lenses~\cite{Michael_Chen_lens}, frequency filters~\cite{FSS2,FSS1}, polarization transformers~\cite{Pol_transformer,Min_CPSS} and electromagnetic cloaks~\cite{michael_S_HMS,paris_cloak,Caloz_cloak,lossless_cloak} have all been successfully demonstrated. Of the various reported metasurface-based devices, those exhibiting omega-bianisotropy have been particularly fascinating, due to their abilities to realize highly sophisticated field transformations with near perfect efficiencies~\cite{local_power_conservation,Suscept_tensor,tretyakov_bianisotropy}. Their robustness stems from the cross-coupling between the electric and magnetic responses, which represents an additional degree of freedom that can be engineered at will.

Much of the research efforts so far in the area of metasurfaces have been devoted to flat planar surfaces. However recently, cylindrical metasurfaces have also generated significant interests due to the potential applications offered by their unique geometries. For instance, the fact that they can completely enclose an object make them ideal for electromagnetic interference reduction and cloaking~\cite{michael_S_HMS,paris_cloak,Caloz_cloak,lossless_cloak}. They have also been used to create electromagnetic illusions~\cite{capolino_illusion,Kwon_Illusion} and conformal antennas~\cite{cylindrical_antenna}. 

Wave scattering by cylindrical metasurfaces can be accurately predicted using integral equations based on tensorial bianisotropic sheet transistion conditions (BSTCs)~\cite{Caloz_MoM,Sandeep_MoM}. Alternatively, one can analyze these surfaces using mode matching~\cite{Mode_Matching}. While the latter technique is more suitable for impedance surfaces, it is possible to generalize it for omega-bianisotropic metasurfaces since they can be constructed from multiple concentric impedance surfaces. By cascading the generalized scattering matrices (GSMs) of its constituent layers, a multilayer omega-bianisotropic metasurface (O-BMS) can be accurately modelled~\cite{transfer_matrix_block}.

Although these approaches provide accurate predictions for the scattered fields of a cylindrical metasurface, they are inherently analysis techniques which do not offer any insights on how to synthesize a surface to perform the desired field transformations. For instance, the cascaded GSMs of impedance surfaces is essentialy a black box that does not reveal how its components should be engineered.

While it is possible to synthesize a cylindrical O-BMS by specifying the desired fields everywhere and solving the BSTC equations to obtain the required surface parameters, the resulting designs will be generally active and/or lossy. These designs are usually undesirable as they require active components and/or accurately controlled ohmic losses to implement in practice. A technique to identify field distributions that correspond to passive and lossless designs has been reported~\cite{lossless_cloak}. However, it is based on numerical optimization of the spatial distribution of electromagnetic fields, which can be difficult.

In this article, we present a mode-matching framework for analysis and synthesis of circular cylindrical O-BMSs inspired by concepts in digital signal processing. The fields everywhere are decomposed into cylindrical modes using the discrete Fourier transform, while the bianisotropic parameters of the metasurface is transformed into their respective spectra. Then, by invoking the BSTCs, we set up systems of algebraic equations which can be easily solved to either model the scattering behaviour of complex O-BMSs, or to synthesize devices that can perform stipulated field transformations. In contrast with previous mode-matching methods, our formulation inherently accounts for the potentially bianisotropic polarizabilities of the metasurface, without needing to decompose it into several constituent parts. This facilitates the derivation of closed-form synthesis equations. Furthermore, the algebraic formulation simplifies the process of identifying passive and lossless field distributions, leading to very practical metasurface designs.

Using the proposed method, we design and investigate several passive and lossless cylindrical O-BMS-based devices. In order to validate the designs, they are each simulated numerically in COMSOL Multiphysics. An electromagnetic illusion metasurface constructed with realistic meta-atoms is designed an verified in Ansys HFSS.

\section{Theory}
\label{sec:theory}
The concepts presented in this paper can be applied to any scalar circular cylindrical omega-bianisotropic metasurfaces. However, some of the steps throughout the analysis and synthesis will require slight problem-specific modifications. In this section, for illustrative purposes, we will investigate the case of an internally excited O-BMS. Generalization to other configurations will be discussed subsequently when they arise.

\begin{figure}[b]
\centering
\includegraphics[width=0.4\textwidth]{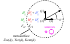}
\caption{Geometric configuration of the internally excited circular cylindrical O-BMS}
\label{fig:geometry}
\end{figure}

The problem under consideration is depicted in Fig.~\ref{fig:geometry}. The metasurface, characterized by its $\phi$-dependent surface electric impedance $Z_{se}(\phi)$, surface magnetic admittance $Y_{sm}(\phi)$ and magnetoelectric coupling coefficient $K_{em}(\phi)$, forms a closed cylindrical cavity of radius $\alpha$. In this paper, we only consider two-dimensional problems which are invariant in the $z$ direction. For sake of simplicity, let us assume the fields are transverse magnetic with respect to the $z$-axis (TM$\mathrm{^z}$) with an implicit time dependency of $e^{j\omega t}$. This means we have $\vec{E}=\hat{z}E_z$. Extension to transverse electric (TE$\mathrm{^z}$) configurations ($\vec{H}=\hat{z}H_z$) is straightforward, and will not be explicated here.

As indicated, the internal and the external regions of the O-BMS have relative permittivities $\epsilon_{r1}$ and $\epsilon_{r2}$ respectively.

The fields enclosed by the cylinder consist of the incident electric and magnetic fields $\{E^i_{||}, H^i_{||}\}$ in addition to the reflected fields $\{E^r_{||}, H^r_{||}\}$. The fields external to the cylinder are just the transmitted fields $\{E^t_{||}, H^t_{||}\}$. The subscript ``$||$'' denotes the transverse components of the fields, which are the components of interest in the present configuration. For TM$\mathrm{^z}$-polarized field distributions, these correspond to the $z$-component of the electric fields and the $\phi$-component of the magnetic fields.

In accordance with the cylindrical coordinate system, the fields inside and outside of the metasurface cavity can be written as series of cylindrical modes described by
\begin{equation}
\label{eqn:external_cylindrical_modes}
\begin{split}
E^{i}_z(\rho,\phi) &= \sum\limits_{p=-\infty}^{\infty}e^i_p\frac{H^{(2)}_p(k_1\rho)}{H^{(2)}_p(k_1\alpha)}e^{-jp\phi},\\
H^{i}_\phi(\rho,\phi) &= \sum\limits_{p=-\infty}^{\infty}e^i_pY^i_{in,p}(\rho)\frac{H^{(2)}_p(k_1\rho)}{H^{(2)}_p(k_1\alpha)}e^{-jp\phi},\\
E^{t}_z(\rho,\phi) &= \sum\limits_{p=-\infty}^{\infty}e^t_p\frac{H^{(2)}_p(k_2\rho)}{H^{(2)}_p(k_2\alpha)}e^{-jp\phi},\\
H^{t}_\phi(\rho,\phi) &= \sum\limits_{p=-\infty}^{\infty}e^t_pY^t_{in,p}(\rho)\frac{H^{(2)}_p(k_2\rho)}{H^{(2)}_p(k_2\alpha)}e^{-jp\phi},\\
E^{r}_z(\rho,\phi) &= \sum\limits_{p=-\infty}^{\infty}e^r_p\frac{J_p(k_1\rho)}{J_p(k_1\alpha)}e^{-jp\phi},\\
H^{r}_\phi(\rho,\phi) &= \sum\limits_{p=-\infty}^{\infty}e^r_pY^r_{in,p}(\rho)\frac{J_p(k_1\rho)}{J_p(k_1\alpha)}e^{-jp\phi},\\
k_{\{1,2\}} &= \sqrt{\epsilon_{r\{1,2\}}}k_o,\\
\end{split}
\end{equation}
where $k_o$ is the free space wave number. $H^{(2)}_{p}(\cdot)$ denotes the $p^{th}$ order Hankel function of second kind and $J_{p}(\cdot)$ denotes the $p^{th}$ order Bessel function. The coefficients $e^{\{i,t,r\}}_p$ are the complex amplitudes of the $p^{th}$ mode of the incident, transmitted and reflected electric fields evaluated at $\rho=\alpha$. 

The $p^{th}$ modes of $E^{\{i,t,r\}}_z$ can be related to the $p^{th}$ modes of $H^{\{i,t,r\}}_\phi$ by the modal wave admittances $Y^{\{i,t,r\}}_{in,p}(\rho)$, where the subscript ``$in,p$" indicates that these are the $p^{th}$ mode admittances for an internally excited metasurface. As will be discussed in Sec.~\ref{sec:results}, the fields of externally excited surface have different modal wave admittances, indicated by their distinct subscripts.

By invoking the linearity of Maxwell's equations and the orthogonality of the cylindrical modal wave functions, the expressions for $Y^{\{i,t,r\}}_{in,p}(\rho)$ can be found to be

\begin{equation}
\label{eqn:wave_admittances}
\begin{split}
Y^i_{in,p}(\rho) &= -\frac{j\sqrt{\epsilon_{r1}}}{\eta_o}\frac{H^{(2)'}_{p}(k_1\rho)}{H^{(2)}_{p}(k_1\rho)},\\
Y^t_{in,p}(\rho) &= -\frac{j\sqrt{\epsilon_{r2}}}{\eta_o}\frac{H^{(2)'}_{p}(k_2\rho)}{H^{(2)}_{p}(k_2\rho)},\\
Y^r_{in,p}(\rho) &= -\frac{j\sqrt{\epsilon_{r1}}}{\eta_o}\frac{J_{p}'(k_1\rho)}{J_{p}(k_1\rho)}.
\end{split}
\end{equation}
The derivatives in (\ref{eqn:wave_admittances}) are evaluated with respect to the entire arguments of $H^{(2)}_p(\cdot)$ and $J_p(\cdot)$.

The exact values of the coefficients $e^{\{i,t,r\}}_p$ can be obtained by performing a Fourier transform on the corresponding fields along a closed circle centered at the origin. In a similar manner, the surface properties $Z_{se}(\phi), Y_{sm}(\phi)$ and $K_{em}(\phi)$ can be decomposed into their constituent Fourier harmonics because they are also $2\pi$-periodic. This idea forms the basis of our analysis and synthesis methods, described in Sec.~\ref{sec:analysis} and Sec.~\ref{sec:synthesis} respectively.

\subsection{Analysis of Internally Excited Circular Cylindrical Omega-Bianisotropic Metasurfaces}
\label{sec:analysis}
We first present a method for evaluating the scattered fields from internally excited circular cylindrical O-BMSs. We begin by discretizing the surface into $N$ equally sized unit cells. The surface properties can be sampled at the centres of these cells, forming $N\times1$ vectors $\bar{Z}_{se}$, $\bar{Y}_{sm}$ and $\bar{K}_{em}$ whose entries are
\begin{equation}
\label{eqn:surface_property_vectors}
\begin{split}
\bar{Z}_{se}[n] &=Z_{se}\big|_{\phi=(n-1)\Delta_\phi},\\
\bar{Y}_{sm}[n] &=Y_{sm}\big|_{\phi=(n-1)\Delta_\phi},\\
\bar{K}_{em}[n] &=K_{em}\big|_{\phi=(n-1)\Delta_\phi},
\end{split}
\end{equation}
where $\Delta_\phi=2\pi/N$ is the azimuthal extent of one unit cell.

We can also sample the tangential fields on the internal and external facets of each unit cell to obtain $N\times1$ field vectors
\begin{equation}
\label{eqn:field_vectors}
\begin{split}
\bar{E}_z^t[n] &= E^t_z\big|_{\rho\to\alpha^+,\phi=(n-1)\Delta_\phi},\\
\bar{E}_z^{\{i,r\}}[n] &= E^{\{i,r\}}_z|_{\rho\to\alpha^-,\phi=(n-1)\Delta_\phi},\\
\bar{H}_\phi^t[n] &= H^t_\phi|_{\rho\to\alpha^+,\phi=(n-1)\Delta_\phi},\\
\bar{H}_\phi^{\{i,r\}}[n] &= H^{\{i,r\}}_\phi|_{\rho\to\alpha^-,\phi=(n-1)\Delta_\phi}.\\
\end{split}
\end{equation}
Alternatively, since all vectors in (\ref{eqn:surface_property_vectors}) and (\ref{eqn:field_vectors}) correspond to quantities that are $2\pi$-periodic in $\phi$, they can be represented in terms of their N-point discrete Fourier transforms (N-DFTs) via multiplications with the N-DFT matrix $\mathbf{W}_N$:
\begin{equation}
\label{eqn:N_DFT}
\begin{split}
\hat{E}^{\{i,r,t\}}_z=\mathbf{W}_N\bar{E}^{\{i,r,t\}}_z,\quad\hat{H}^{\{i,r,t\}}_z=\mathbf{W}_N\bar{H}^{\{i,r,t\}}_z,\\
\hat{Z}_{se} = \mathbf{W}_N\bar{Z}_{se},\quad\hat{Y}_{sm} =\mathbf{W}_N\bar{Y}_{sm},\quad\hat{K}_{em} = \mathbf{W}_N\bar{K}_{em}.
\end{split}
\end{equation}
$\mathbf{W}_N$ is the N-DFT matrix whose elements are
\begin{equation}
\label{eqn:DFT_matrix}
\begin{split}
\mathbf{W}_N[n][m]&=\frac{1}{N}e^{j\frac{2\pi}{N}n^\star\big(m-1\big)},\\
n^\star&\triangleq n-\frac{N+1}{2}.
\end{split}
\end{equation}
We refer to $\hat{E}^{\{i,r,t\}}_z$ and $\hat{H}^{\{i,r,t\}}_\phi$ as the modal field vectors, since their entries represent the amplitudes of the constituent cylindrical modes of $E^{\{i,r,t\}}_z$ and $H^{\{i,r,t\}}_\phi$, evaluated at $\rho=\alpha$. On the other hand, $\hat{Z}_{se}$, $\hat{Y}_{sm}$ and $\hat{K}_{em}$ are the surface property spectra whose entries represent the amplitudes of the Fourier harmonics constituting $\bar{Z}_{se}$, $\bar{Y}_{sm}$ and $\bar{K}_{em}$.

The rows of $\mathbf{W}_N$, as defined by (\ref{eqn:DFT_matrix}), are arranged such that the center entries of the modal field vectors and the surface property spectra correspond to $p=0$ expansion term. The first and the last entries correspond to the $p=(1-N)/2\triangleq p_-$ and the $p=(N-1)/2\triangleq p_+$ terms respectively. 

One advantage of  the modal analysis presented herein is that $\hat{E}^{\{i,r,t\}}_z$  can be algebraically related to $\hat{H}^{\{i,r,t\}}_\phi$  with the help of (\ref{eqn:wave_admittances}). More specifically, we can write
\begin{equation}
\label{eqn:elimination_H}
\begin{split}
\hat{H}^{\{i,t,r\}}_\phi &=\mathbf{Y}_{in}^{\{i,t,r\}}\hat{E}^{\{i,t,r\}}_z,\\
\mathbf{Y}_{in}^{\{i,t,r\}} &=\mathrm{diag}\begin{bmatrix}  	Y^{\{i,t,r\}}_{in,p_-}(\alpha)  &
																					\cdots &
																					Y^{\{i,t,r\}}_{in,p_+}(\alpha) \\
																					\end{bmatrix}^T,
\end{split}
\end{equation}
where $\mathbf{Y}_{in}^{\{i,t,r\}}$ are the diagonal modal admittance matrices for the incident, transmitted and reflected fields. They can be used to eliminate all magnetic field quantities from the analysis, thereby reducing the number of unknowns. Due to the different wave admittance expressions, other configurations such as externally excited surfaces may require slightly different modal admittance matrices.

The original goal of this section is to predict the transmitted and reflected fields, given the incident fields and the O-BMS surface parameters. To this end, we now relate the field vectors on either side of the metasurface by invoking the BSTCs at each of the $N$ unit cells~\cite{local_power_conservation}. Doing so results in the system of equations
\begin{equation}
\label{eqn:discrete_BSTC}
\begin{split}
\frac{1}{2}(\bar{E}^t_{z}+\bar{E}^i_{z}+\bar{E}^r_{z})= &\bar{Z}_{se}\odot(\bar{H}^t_{\phi}-\bar{H}^i_{\phi}-\bar{H}^r_{\phi})\\
															  &-\bar{K}_{em}\odot(\bar{E}^t_{z}-\bar{E}^i_{z}-\bar{E}^r_{z}),\\
\frac{1}{2}(\bar{H}^t_{\phi}+\bar{H}^i_{\phi}+\bar{H}^r_{\phi}) = &\bar{Y}_{sm}\odot(\bar{E}^t_{z}-\bar{E}^i_{z}-\bar{E}^r_{z})\\
                                                                  &+\bar{K}_{em}\odot(\bar{H}^t_{\phi}-\bar{H}^i_{\phi}-\bar{H}^r_{\phi}),\\
\end{split}
\end{equation}
where $\odot$ denotes element-wise multiplication.

In order to take advantage of the simplifications offered by modal analysis, we transform (\ref{eqn:discrete_BSTC}) using the circular convolution theorem for N-DFT, which says
\begin{equation}
\label{eqn:circular_convolution_theorem}
\begin{split}
\mathbf{W}_N(\bar{U}\odot\bar{V})&=\hat{U}\circledast_N\hat{V}=\mathbf{\hat{U}}\hat{V}.\\
\end{split}															
\end{equation}
The operation $\circledast_N$ denotes modulo-N circular convolution. As alluded to by the second equality in (\ref{eqn:circular_convolution_theorem}), we can implement this operation as a matrix multiplication by transforming the first operand $\hat{U}$ into the $N\times N$ circulant matrix $\mathbf{\hat{U}}$ given by

\begin{equation}
\mathbf{\hat{U}}=\mathbf{W}_N\mathrm{diag}(\mathbf{W}^{-1}_N\hat{U})\mathbf{W}^{-1}_N
\end{equation}

Applying N-DFT to both sides of (\ref{eqn:discrete_BSTC}) and applying (\ref{eqn:elimination_H}) and (\ref{eqn:circular_convolution_theorem}), we obtain a new system of equations
\begin{equation}
\label{eqn:spectral_BSTC_cylindrical}
\begin{split}
&\scaleto{\frac{1}{2}(\hat{E}^t_{z}+\hat{E}^i_{z}+\hat{E}^r_{z})}{20pt} \\
= &\mathbf{\hat{Z}}(\mathbf{Y}_{in}^t\hat{E}^t_{z}-\mathbf{Y}_{in}^i\hat{E}^i_{z}-\mathbf{Y}_{in}^r\hat{E}^r_{z})-\mathbf{\hat{K}}(\hat{E}^t_{z}-\hat{E}^i_{z}-\hat{E}^r_{z}),\\
&\frac{1}{2}(\mathbf{Y}_{in}^t\hat{E}^t_{z}+\mathbf{Y}_{in}^i\hat{E}^i_{z}+\mathbf{Y}_{in}^r\hat{E}^r_{z})\\
 =&\mathbf{\hat{Y}}(\hat{E}^t_{z}-\hat{E}^i_{z}-\hat{E}^r_{z})+\mathbf{\hat{K}}(\mathbf{Y}_{in}^t\hat{E}^t_{z}-\mathbf{Y}_{in}^i\hat{E}^i_{z}-\mathbf{Y}_{in}^r\hat{E}^r_{z}),\\
\end{split}
\end{equation}
where $\mathbf{\hat{Z}}$, $\mathbf{\hat{Y}}$ and $\mathbf{\hat{K}}$ are $N\times N$ circulant matrices formed by $\hat{Z}_{se}$, $\hat{Y}_{sm}$ and $\hat{K}_{em}$ respectively. This set of equations can be easily rearranged to solve for the modal transmission matrix $\mathbf{\hat{T}}_{in}$ and the modal reflection matrix $\mathbf{\hat{R}}_{in}$; they relate $\hat{E}^t_z$ and $\hat{E}^r_z$ to $\hat{E}^i_z$ according to
\begin{equation}
\label{eqn:modal_transmission_reflection}
\hat{E}^t_z=\mathbf{\hat{T}}_{in}\hat{E}^i_z,\quad \hat{E}^r_z=\mathbf{\hat{R}}_{in}\hat{E}^i_z.
\end{equation}
The exact solution for the two matrices are

\begin{equation}
\label{eqn:T_matrix}
\begin{split}
\mathbf{\hat{T}}_{in} = &\mathbf{\hat{t}}_{in,b}^{-1}\mathbf{\hat{t}}_{in,a},\\
\mathbf{\hat{t}}_{in,a} = &\scaleto{\bigg(\frac{1}{2}\mathbf{I}-\mathbf{\hat{K}}+\mathbf{\hat{Z}}\mathbf{Y}_{in}^r\bigg)^{-1}\bigg(\frac{1}{2}\mathbf{I}-\mathbf{\hat{K}}+\mathbf{\hat{Z}}\mathbf{Y}_{in}^i\bigg)}{24pt}\\
 								&\scaleto{-\bigg(\frac{1}{2}\mathbf{Y}_{in}^r+\mathbf{\hat{Y}}+\mathbf{\hat{K}}\mathbf{Y}_{in}^r\bigg)^{-1}\bigg(\frac{1}{2}\mathbf{Y}_{in}^i+\mathbf{\hat{Y}}+\mathbf{\hat{K}}\mathbf{Y}_{in}^i\bigg),}{24pt}\\
\mathbf{\hat{t}}_{in,b} = &\scaleto{\bigg(\frac{1}{2}\mathbf{Y}_{in}^r+\mathbf{\hat{Y}}+\mathbf{\hat{K}}\mathbf{Y}_{in}^r\bigg)^{-1}\bigg(\frac{1}{2}\mathbf{Y}_{in}^t-\mathbf{\hat{Y}}-\mathbf{\hat{K}}\mathbf{Y}_{in}^t\bigg)}{24pt}\\
							   &\scaleto{-\bigg(\frac{1}{2}\mathbf{I}-\mathbf{\hat{K}}+\mathbf{\hat{Z}}\mathbf{Y}_{in}^r\bigg)^{-1}\bigg(\frac{1}{2}\mathbf{I}+\mathbf{\hat{K}}-\mathbf{\hat{Z}}\mathbf{Y}_{in}^t\bigg),}{24pt}\\
\mathbf{\hat{R}}_{in} = &\mathbf{\hat{r}}_{in,b}^{-1}\mathbf{\hat{r}}_{in,a},\\
\mathbf{\hat{r}}_{in,a} = &\scaleto{\bigg(\frac{1}{2}\mathbf{I}+\mathbf{\hat{K}}-\mathbf{\hat{Z}}\mathbf{Y}_{in}^t\bigg)^{-1}\bigg(\frac{1}{2}\mathbf{I}-\mathbf{\hat{K}}+\mathbf{\hat{Z}}\mathbf{Y}_{in}^i\bigg)}{24pt}\\
								&\scaleto{-\bigg(\frac{1}{2}\mathbf{Y}_{in}^t-\mathbf{\hat{Y}}-\mathbf{\hat{K}}\mathbf{Y}_{in}^t\bigg)^{-1}\bigg(\frac{1}{2}\mathbf{Y}_{in}^i+\mathbf{\hat{Y}}+\mathbf{\hat{K}}\mathbf{Y}_{in}^i\bigg),}{24pt}\\
\mathbf{\hat{r}}_{in,b} = &\scaleto{\bigg(\frac{1}{2}\mathbf{Y}_{in}^t-\mathbf{\hat{Y}}-\mathbf{\hat{K}}\mathbf{Y}_{in}^t\bigg)^{-1}\bigg(\frac{1}{2}\mathbf{Y}_{in}^r+\mathbf{\hat{Y}}+\mathbf{\hat{K}}\mathbf{Y}_{in}^r\bigg)}{24pt}\\
								&\scaleto{-\bigg(\frac{1}{2}\mathbf{I}+\mathbf{\hat{K}}-\mathbf{\hat{Z}}\mathbf{Y}_{in}^t\bigg)^{-1}\bigg(\frac{1}{2}\mathbf{I}-\mathbf{\hat{K}}+\mathbf{\hat{Z}}\mathbf{Y}_{in}^r\bigg).}{24pt}\\
\end{split}
\end{equation}
Here, $\mathbf{I}$ is the $N\times N$ identity matrix. Since $\hat{H}_\phi^t$ and $\hat{H}_\phi^r$ can be found using (\ref{eqn:elimination_H}), we have completed our analysis.

\subsection{Synthesis of Internally Excited Circular Cylindrical Omega-Bianisotropic Metasurfaces}
\label{sec:synthesis}
Within the proposed DFT framework, it is also possible to synthesize the required surface properties for realizing a stipulated field transformation.

While the analysis presented in Sec.~\ref{sec:analysis} applies to general O-BMSs that can contain power gain and/or loss, we will devote our effort henceforth to the synthesis of passive and lossless metasurfaces. This is because they are much easier to implement in practice, requiring only reactive components which can be realized with etched patterns on printed circuit boards (PCBs). Often, the undesirable losses associated with realistic PCBs are not pronounced enough to significantly corrupt the functionalities of a fabricated device, even if losslessness is assumed throughout the entire design process~\cite{Chen_refraction}.

Previously, it was shown that a sufficient condition for an O-BMS to be passive and lossless is that the field transformation being performed satisfies local power conservation (LPC)~\cite{local_power_conservation}, which can be described by the equation
\begin{equation}
\label{eqn:spatial_LPC}
\mathrm{Re}\left\{(\bar{E}_z^i+\bar{E}_z^r)\odot(\bar{H}_\phi^i+\bar{H}_\phi^r)^*\right\}=\mathrm{Re}\left\{\bar{E}_z^t\odot\bar{H}_\phi^{t^*}\right\},
\end{equation}
where $(\cdot)^*$ denotes complex conjugation. Although composite metasurface systems which satisfy global power conservation represent possible alternatives~\cite{Ayman_nonlocal}, they are much harder to design and implement. Thus, we will strictly focus on surfaces that satisfy LPC. From (\ref{eqn:spatial_LPC}), it can be inferred that a passive and lossless transmissive scalar metasurface, which is one that produces a prescribed transmitted field distribution from a known incident field, should generate some amount of parasitic reflected fields (termed ``auxiliary reflection"). Similarly, a passive and lossless reflective scalar metasurface will require some auxiliary transmission. Naturally, depending on the type of metasurface to be synthesized, we would first need to solve for either $\{\bar{E}^r,\bar{H}^r\}$ or $\{\bar{E}^t,\bar{H}^t\}$. To do this directly using (\ref{eqn:spatial_LPC}) can be challenging. One would need to solve a system of coupled nonlinear differential equations since the unknown auxiliary electric fields and auxiliary magnetic fields are related through Maxwell's equations. In this work, we use (\ref{eqn:elimination_H}) to eliminate $\bar{H}^{\{i,t,r\}}_\phi$ from the equation, yielding a non-linear algebraic equation which can be solved numerically with ease. For transmissive metasurfaces, where $\bar{E}^r_z$ or $\hat{E}^r_z$ is the unknown, we have
\begin{equation}
\label{eqn:modified_LPC}
\begin{split}
&\mathrm{Re}\{(\bar{E}_z^i+\mathbf{W}_N^{-1}\hat{E}_z^r)\odot(\bar{H}_\phi^i+\mathbf{W}_N^{-1}\mathbf{Y}_{in}^r\hat{E}_z^r)^*\}\\
=&\mathrm{Re}\{\bar{E}_z^t\odot\bar{H}_\phi^{t^*}\}.
\end{split}
\end{equation}
For reflective metasurfaces, where $\bar{E}^t_z$ or $\hat{E}^t_z$ is the unknown, the equation to solve is
\begin{equation}
\label{eqn:modified_LPC_2}
\begin{split}
&\mathrm{Re}\{(\bar{E}_z^i+\bar{E}_z^r)\odot(\bar{H}_\phi^i+\bar{E}_z^r)^*\}\\
=&\mathrm{Re}\{(\mathbf{W}_N^{-1}\hat{E}_z^t)\odot(\mathbf{W}_N^{-1}\mathbf{Y}_{in}^t\hat{E}_z^{t})^*\}.
\end{split}
\end{equation}

Solving either (\ref{eqn:modified_LPC}) or (\ref{eqn:modified_LPC_2}) gives the complete field distributions everywhere. We can proceed by assessing the required $\{\bar{Z}_{se},\bar{Y}_{sm},\bar{K}_{em}\}$ which would support these field distributions. 

In Sec.~\ref{sec:analysis}, we transformed the known O-BMS parameters into circulant matrices and solved for the unknown field modal vectors. Here, we can perform the inverse procedure by transforming the known field distributions into circulant matrices and solving for the unknown surface property spectra. This leads to the equations
\begin{equation}
\label{eqn:Spectral_BSTC2}
\begin{split}
\frac{1}{2}(\hat{E}^t_z+\hat{E}^i_z+\hat{E}^r_z)=\hat{E}_{av}&=\mathbf{\Delta}_{\hat{H},in}\hat{Z}_{se} -\mathbf{\Delta}_{\hat{E},in}\hat{K}_{em},\\
\frac{1}{2}(\hat{H}^t_\phi+\hat{H}^i_\phi+\hat{H}^r_\phi)=\hat{H}_{av}&=\mathbf{\Delta}_{\hat{E},in}\hat{Y}_{sm} +\mathbf{\Delta}_{\hat{H},in}\hat{K}_{em},\\
\end{split}
\end{equation}
where $\mathbf{\Delta}_{\hat{E},in}$ and $\mathbf{\Delta}_{\hat{H},in}$ are the $N\times N$ field discontinuity circulant matrices formed using the vectors $(\hat{E}^t_z-\hat{E}^r_z-\hat{E}^i_z)$ and $(\hat{H}^t_\phi-\hat{H}^r_\phi-\hat{H}^i_\phi)$ respectively.

An O-BMS satisfying LPC will have imaginary $\bar{Z}_{se}$ and $\bar{Y}_{sm}$, as well as real $\bar{K}_{em}$~\cite{passive_lossless_parameter}. This allows us to reduce the number of unknowns in (\ref{eqn:Spectral_BSTC2}) by invoking the conjugate symmetry properties for the DFT of real and imaginary signals:
\begin{equation}
\begin{split}
\label{eqn:DFT_symmetry}
\mathrm{Re}\{\bar{V}\}=0 &\implies \mathbf{F}\hat{V}=-\hat{V}^*,\\
\mathrm{Im}\{\bar{V}\}=0 &\implies \mathbf{F}\hat{V}=\hat{V}^*,\\
\end{split}
\end{equation}
where $\mathbf{F}$ is the $N\times N$ reversal matrix. Using the fact that $\mathbf{F}\mathbf{F}=\mathbf{I}$, we can obtain another system of equation from~(\ref{eqn:Spectral_BSTC2}) as follows:
\begin{equation}
\label{eqn:Spectral_BSTC2_2}
\begin{split}
\hat{E}_{av}&=-\mathbf{\Delta}_{\hat{H},in}\mathbf{F}\hat{Z}^*_{se} -\mathbf{\Delta}_{\hat{E},in}\mathbf{F}\hat{K}^*_{em},\\
\hat{H}_{av}&=-\mathbf{\Delta}_{\hat{E},in}\mathbf{F}\hat{Y}^*_{sm} +\mathbf{\Delta}_{\hat{H},in}\mathbf{F}\hat{K}^*_{em}.\\
\end{split}
\end{equation}
Combining (\ref{eqn:Spectral_BSTC2}) with the complex conjugate of (\ref{eqn:Spectral_BSTC2_2}), we have sufficiently many independent equations to solve for the unknown surface property spectra as

\begin{equation}
\label{eqn:Spectral_BSTC2_solution}
\begin{split}
\hat{K}_{em} = &\mathbf{k}_a^{-1}\bar{k}_b,\\
\mathbf{k}_a = &-\big(\mathbf{\Delta}_{\hat{H},in}^*\mathbf{F}\big)^{-1}(\mathbf{\Delta}_{\hat{E},in}^*\mathbf{F})-\mathbf{\Delta}_{\hat{H},in}^{-1}\mathbf{\Delta}_{\hat{E},in},\\
\bar{k}_b = &\big(\mathbf{\Delta}_{\hat{H},in}^*\mathbf{F}\big)^{-1}\hat{E}_{av}^*+\mathbf{\Delta}_{\hat{H},in}^{-1}\hat{E}_{av},\\
\hat{Z}_{se} = & \mathbf{\Delta}_{\hat{H},in}^{-1}\hat{E}_{av}+\mathbf{\Delta}_{\hat{H},in}^{-1}\mathbf{\Delta}_{\hat{E},in}\hat{K}_{em},\\
\hat{Y}_{sm} = &\mathbf{\Delta}_{\hat{E},in}^{-1}\hat{H}_{av}-\mathbf{\Delta}_{\hat{E},in}^{-1}\mathbf{\Delta}_{\hat{H},in}\hat{K}_{em}.\\
\end{split}
\end{equation}
This concludes the synthesis procedure of internally excited passive and lossless O-BMS.

\subsection{Multilayer Implementations}
\label{sec:transfer_matrix}
\begin{figure}[b]
\centering
\includegraphics[width=0.49\textwidth]{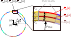}
\caption{Triple impedance layer unit cell topology used to realize the theoretically derived O-BMS properties.}
\label{fig:impedance_layer_geometry}
\end{figure}

Having obtained the theoretical O-BMS surface parameters in Sec.~\ref{sec:synthesis}, we now consider a physical unit cell topology that is suitable for practical realization of the derived properties. For planar metasurfaces, unit cells consisting of three parallel electric impedance sheets separated by dielectric substrates have often been used~\cite{Joseph}. It was shown that the unit cell can exhibit omega-bianisotropic response if its three constituent layers are asymmetric with respect to the middle one (i.e. the top and the bottom layers have different impedances). This topology is usually preferred because it is highly compatible with standard PCB fabrication technologies. The three impedance sheets can be easily realized using three layers of etched conductive patterns~\cite{Chen1, Abdo}. In this section, we consider an analogous curved triple impedance layer topology, depicted in Fig.~\ref{fig:impedance_layer_geometry}. 

Each unit cell of the surface is assigned a set of inner, middle and outer impedance values; they are denoted as $\bar{Z}_i[n]$, $\bar{Z}_m[n]$ and $\bar{Z}_o[n]$ respectively, for the $n^{th}$ cell. The three impedance layers are supported by two cylindrical dielectric shells with thickness $t$ and relative permittivity $\epsilon_r$. The inner surface impedances $\bar{Z}_i$ reside on the cylindrical surface $\rho=\alpha$.

For planar O-BMSs, equivalent transmission-line models have frequently been used to aid the design of the multilayer unit cells~\cite{Paul_AHMS1}. It is hard to directly adopt this approach here because the wave impedances for cylindrical waves are dependent on the radial coordinate $\rho$. Thus, they vary throughout the longitudinal extent of each unit cell, making it difficult to construct a simple equivalent transmission-line circuit. Instead, we employ a generalization of the ABCD matrix approach previously used to design a mode-converting cylindrical O-BMS~\cite{angular_momentum}.

Before proceeding, we note that the ABCD matrix approach assumes local periodicity. That is, each unit cell is analyzed and designed as if it resides within a homogeneous surface. An accompanying assumption is that only the $p=0$ mode is propagating between the multiple impedance layers. In a realistic metasurface, due to the $\phi$-dependent impedances, local periodicity is not satisfied and higher order modes can propagate within the extent of each unit cell. This can make the ABCD matrix approach inaccurate, especially if the dielectric layers are thick~\cite{FP_HMS}. To alleviate this phenomenon, we employ perfect conducting baffles to shield each unit cell from its neighbours. This approach is inspired by a previous work which used baffles to decouple the adjacent unit cells in a planar bianisotropic metasurface~\cite{Paul_AHMS2}. As discussed therein, the baffles act as an array of small waveguides which cut off all but the fundamental mode  locally within each unit cell. For the TM$\mathrm{^z}$ configuration discussed in this study, perfect magnetic conductor (PMC) baffles are used, as seen in Fig.~\ref{fig:impedance_layer_geometry}. For TE$\mathrm{^z}$ configurations, perfect electric conductor (PEC) baffles are required. While the PEC baffles are more practical since they can be fabricated using plated via fences, PMC baffles will mainly serve as a numerical tools to aid the validation of theoretically synthesized metasurfaces.

Assuming that the ABCD matrix approach is valid, we can apply it to each unit cell individually. The ABCD matrix $\mathbf{M}$ between two concentric cylindrical surfaces with radii $\rho_1$ and $\rho_2$ relates the \textit{total} electric and magnetic fields on those surfaces according to

\begin{equation}
\label{eqn:transfer_matrix_definition}
\begin{bmatrix}
E_z(\rho_1)\\
H_\phi(\rho_1)
\end{bmatrix}=\mathbf{M}\begin{bmatrix}
E_z(\rho_2)\\
H_\phi(\rho_2)
\end{bmatrix}=\begin{bmatrix}
A & B\\C & D\end{bmatrix}\begin{bmatrix}
E_z(\rho_2)\\
H_\phi(\rho_2)
\end{bmatrix}.
\end{equation}

For instance, the ABCD matrix relating the fields on either side of a cylindrical sheet with electric impedance $Z$ is
\begin{equation}
\mathbf{M}_Z=\begin{bmatrix}
								1&0\\
								-\frac{1}{Z}&1\\
								\end{bmatrix}.
\end{equation}
This can be easily derived from the BSTC equations.

We can also find the ABCD matrix for the dielectric substrates~\cite{transfer_matrix_block}. If the inner and outer radii of the substrate are $\rho_i$ and $\rho_o$ respectively, and the relative permittivity is $\epsilon_r$, then the matrix is given by
\begin{equation}
\label{eqn:transfer_matrix_cylinder}
\begin{split}
\mathbf{M}_{sub}(\rho_i,\rho_o)&=\begin{bmatrix} A(\rho_i,\rho_o) & B(\rho_i,\rho_o)\\
																						 C(\rho_i,\rho_o)&D(\rho_i,\rho_o)\\\end{bmatrix}=\mathbf{m}_i\mathbf{m}_o^{-1},\\
\mathbf{m}_{\{i,o\}}[1][1] & = H^{(2)}_0\left(\sqrt{\epsilon_r}k_o\rho_{\{i,o\}}\right),\\
\mathbf{m}_{\{i,o\}}[1][2] & = J_0\left(\sqrt{\epsilon_r}k_o\rho_{\{i,o\}}\right),\\
\mathbf{m}_{\{i,o\}}[2][1] & = Y_{sub}^+\left(\rho_{\{i,o\}}\right)H^{(2)}_0\left(\sqrt{\epsilon_r}k_o\rho_{\{i,o\}}\right),\\
\mathbf{m}_{\{i,o\}}[2][2] & = Y^-_{sub}\left(\rho_{\{i,o\}}\right)J_0\left(\sqrt{\epsilon_r}k_o\rho_{\{i,o\}}\right),\\
\end{split}
\end{equation}
where
\begin{equation}
\begin{split}
Y_{sub}^+(\rho) & = -\frac{j\sqrt{\epsilon_{r}}}{\eta_o}\frac{H^{(2)'}_{0}(\sqrt{\epsilon_{r}}k_o\rho)}{H^{(2)}_{0}(\sqrt{\epsilon_{r}}k_o\rho)},\\
Y_{sub}^-(\rho) & = -\frac{j\sqrt{\epsilon_{r}}}{\eta_o}\frac{J_{0}'(\sqrt{\epsilon_{r}}k_o\rho)}{J_{0}(\sqrt{\epsilon_{r}}k_o\rho)}.
\end{split}
\end{equation}

From Fig.~\ref{fig:impedance_layer_geometry}, we can see that the matrices \mbox{$\mathbf{M}_{sub}(\alpha,\alpha+t)$} and $\mathbf{M}_{sub}(\alpha+t,\alpha+2t)$ model the inner and the outer dielectric substrates respectively. Cascading the matrices for the three impedance layers as well as those for the two dielectric substrates in the appropriate order, we obtain the overall ABCD matrix for a unit cell
\begin{equation}
\label{eqn:cascaded_transfer_matrix}
\mathbf{M}=\mathbf{M}_{Z_{i}}\mathbf{M}_{sub}(\alpha,\alpha+t)\mathbf{M}_{Z_{m}}\mathbf{M}_{sub}(\alpha+t,\alpha+2t)\mathbf{M}_{Z_{o}},
\end{equation}
where $Z_{i}$, $Z_{m}$, $Z_{o}$ are the inner, middle and outer impedances for that cell respectively.

Assuming we have already calculated the O-BMS parameters $\{Z_{se}, Y_{sm}, K_{em}\}$ for the unit cell following the procedures in Sec.~\ref{sec:synthesis}, we can use (\ref{eqn:discrete_BSTC})~and~(\ref{eqn:transfer_matrix_definition}) to find the equivalent ABCD matrix parameters of that cell as
\begin{equation}
\label{eqn:ABCD_ZYK}
\begin{split}
A &= \frac{4K_{em}^2+4Y_{sm}Z_{se}+4K_{em}+1}{4K_{em}^2+4Y_{sm}Z_{se}-1},\\
B &= \frac{-4Z_{se}}{4K_{em}^2+4Y_{sm}Z_{se}-1},\\
C &= \frac{-4Y_{sm}}{4K_{em}^2+4Y_{sm}Z_{se}-1},\\
D &= \frac{4K_{em}^2+4Y_{sm}Z_{se}-4K_{em}+1}{4K_{em}^2+4Y_{sm}Z_{se}-1}.\\
\end{split}
\end{equation}

If we denote the elements of dielectric substrate ABCD matrices as
\begin{equation}
\label{eqn:dielectric_matrix_elements}
\begin{split}
\mathbf{M}_{sub}(\alpha+t,\alpha+2t) &= \begin{bmatrix} a_1&b_1\\c_1&d_1\end{bmatrix},\\
\mathbf{M}_{sub}(\alpha,\alpha+t) &= \begin{bmatrix} a_2&b_2\\c_2&d_2\end{bmatrix},\\
\end{split}
\end{equation}
then the required $Z_{i}$, $Z_{m}$ and $Z_{o}$ can be found to be
\begin{equation}
\label{eqn:required_impedance_sheets}
\begin{split}
Z_i &= \frac{-b_2B}{b_2D-d_2B+a_2b_1d_2-b_1b_2c_2},\\
Z_m &= \frac{b_1b_2}{a_2b_1-B+b_2d_1},\\
Z_o &= \frac{b_1B}{a_1B - b_1A - a_1b_2d_1 + b_1b_2c_1}.
\end{split}
\end{equation}

Using (\ref{eqn:ABCD_ZYK}) and (\ref{eqn:required_impedance_sheets}) on each of the $N$ unit cells gives the final theoretical implementation of our synthesized O-BMS. 

\subsection{Comments on Other Metasurface Configurations}
\label{sec:alt_config}
As mentioned earlier, some of the steps in the analysis and synthesis procedures need to be modified according to the problem geometry.  This requirement stems from the fact that the modal expansions of the electric and magnetic fields are problem-specific. Furthermore, the BSTC equations (\ref{eqn:discrete_BSTC}) are different depending on the location of the source (internal or external). This in turn leads to different analysis (\ref{eqn:T_matrix}) and synthesis (\ref{eqn:Spectral_BSTC2_solution}) equations. On the other hand, the ABCD matrix formalism presented in Sec.~\ref{sec:transfer_matrix} does not assume anything about the source location, and thus can be used for any problem. As we go over detailed design examples in Sec.~\ref{sec:results}, it will become clear that the overall DFT framework is readily generalizable.

\section{Results and Discussions}
\label{sec:results}
In this section, we design several passive and lossless O-BMSs using our proposed method and validate them numerically using finite element simulations in COMSOL Multiphysics. Without loss of generality, we assume henceforth that the external region is comprised of air ($\epsilon_{r2}=1$).
\subsection{Electromagnetic Illusion}
\label{sec:EM_Illusion}
In the first example, we design a cylindrical O-BMS which transforms the fields radiated by an electric line source located at the origin to those from a displaced line source. Previously, metamaterials (MTMs) based on transformation optics (TO) have been leveraged to achieve this effect~\cite{TO_illusion}. However, the practicality of TO-MTMs is significantly hindered by their bulkiness and complexity. Here, we attempt to achieve the same electromagnetic illusion using a metasurface with deeply subwavelength thickness, which is much easier to fabricate and deploy. Although this type of device has been widely reported~\cite{Caloz_MoM, Sandeep_MoM,capolino_illusion}, passive lossless implementations leveraging omega-bianisotropy have not been demonstrated thus far to the best of our knowledge. 

To design the illusion metasurface, we first state the incident electric field modal vector as
\begin{equation}
\label{eq:ex1_inc} 
\hat{E}_z^i[n]=e^i_0\delta_{n^\star,0},
\end{equation}
where $n^\star$ is as defined in (\ref{eqn:DFT_matrix}) and $\delta_{i,j}$ is the Kronecker delta. The complex amplitude $e^i_0$ can be an arbitrarily chosen. There is only one non-zero entry in $\hat{E}_z^i $ because an electric line source located at the origin can only excite the 0$^{th}$ cylindrical mode.

The transmitted modal vector can be obtained using the addition theorem, which describes a Hankel function centred at $(\rho,\phi)=(\rho',\phi')$ in terms of a summation of cylindrical modes centred at the origin~\cite{balanis}:

\begin{subnumcases}
{\label{eqn:hankel_addition} \scaleto{H^{(2)}_0(k_o|\vec{\rho}-\pvec{\rho}'|)=}{10pt}}
   \scaleto{\sum\limits_{p=-\infty}^{\infty}J_p(k_o\rho')H_p^{(2)}(k_o\rho)e^{jp(\phi-\phi')}}{24pt} & $\scaleto{\rho\geq \rho'}{8pt}$ \label{eqn:hankel_internal_source}
   \\
   \scaleto{\sum\limits_{p=-\infty}^{\infty}H_p^{(2)}(k_o\rho')J_p(k_o\rho)e^{jp(\phi-\phi')}}{24pt} & $\scaleto{\rho\leq \rho'}{8pt}$ \label{eqn:hankel_external_source}
\end{subnumcases}

\begin{figure}[hb!]
\centering  
\subfigure[]{\includegraphics[width=0.98\linewidth]{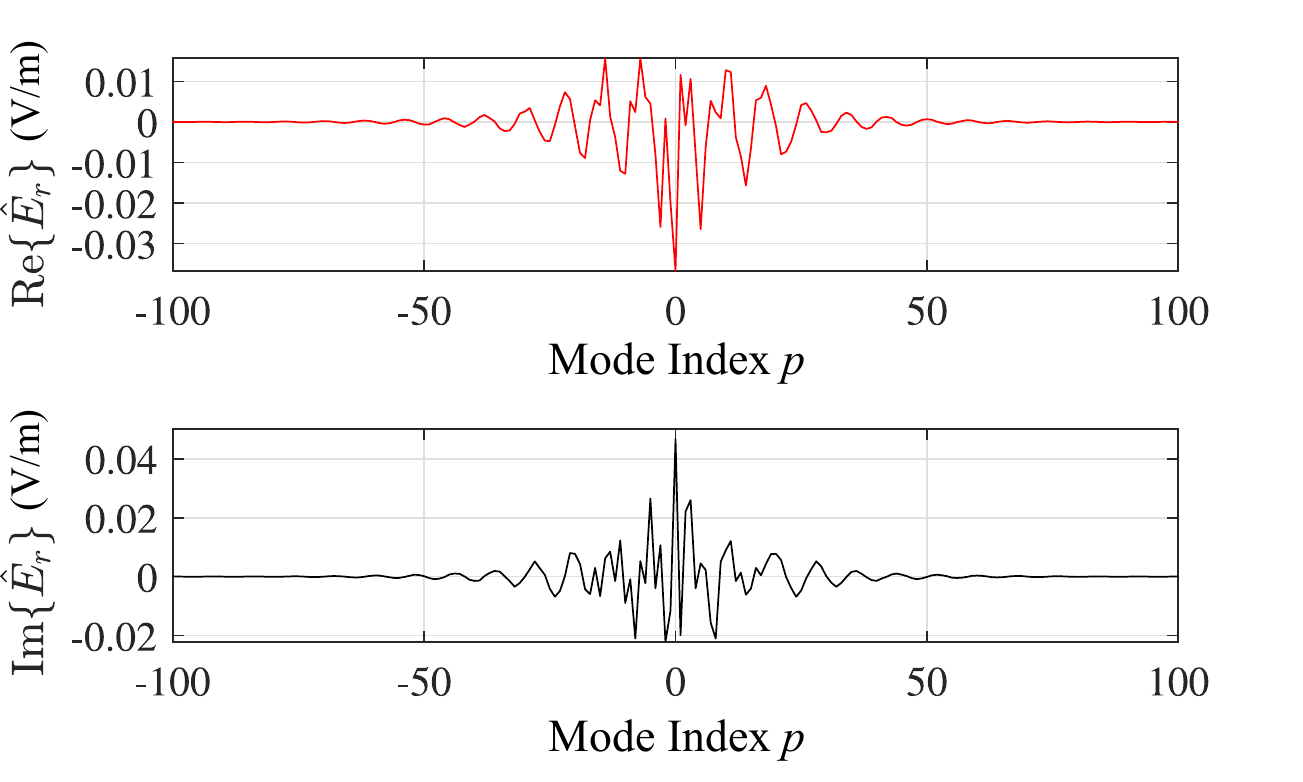}}
\subfigure[]{\includegraphics[width=0.98\linewidth]{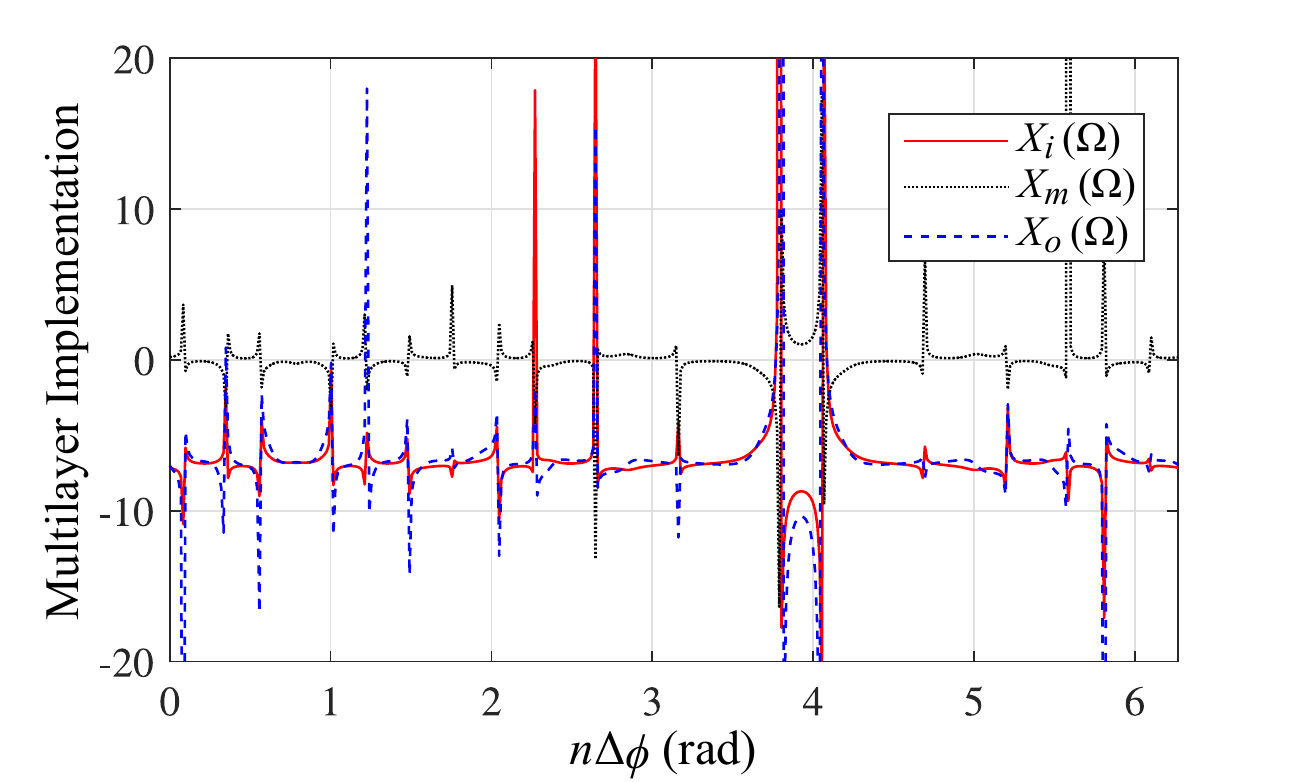}}
\caption{(a) The solved auxiliary reflected fields for the illusion O-BMS. Higher-order modes with near-zero amplitudes are omitted. (b) The reactance values for the multilayer implementation of the illusion O-BMS.}
\label{fig:ex1}
\end{figure}

For the case of a virtual source located inside the O-BMS cavity ($\rho'<\alpha$), we can obtain $\hat{E}^t_z$ using (\ref{eqn:hankel_internal_source}) by setting $\rho=\alpha$:

\begin{equation}
\label{eqn:ex1_trans}
\hat{E}_z^t[n] = e^t_0\frac{J_{n^\star}(k_o\rho')H^{(2)}_{n^\star}(k_o\alpha)}{H^{(2)}_0(k_o\alpha)}e^{jn^\star\phi'}.
\end{equation}
The coefficient $e^t_0$ is the amplitude of the virtual source field measured at a distance $\alpha$ away from $\pvec{\rho}'$; it is set to be equal to $e^i_0$ in this study. As an example, we design an illusion metasurface with the specifications listed in Table~\ref{tab:ex1_spec}. Since this is a transmissive metasurface, we solve for the required auxiliary $\hat{E}^r_z$ using (\ref{eqn:modified_LPC}). The numerically obtained solution is shown in in Fig.~\ref{fig:ex1}(a). Using the complete field distributions $\hat{E}^{\{i,t,r\}}_z$, we obtain the multilayer implementation for the O-BMS described by Fig.~\ref{fig:ex1}(b). Notably, we only plot the reactances of each layer, since the real parts of the impedances are identically zero. This indicates the synthesized device is truly passive and lossless.

\begin{table}[t]
\begin{ruledtabular}
\caption{Specification for the passive lossless illusion O-BMS}
\label{tab:ex1_spec}
\begin{tabular}{ccccccccc}
  $f$ (GHz) & $N$ & $\alpha$ (m) & $\rho'$ & $\phi'$ (rad) & $\epsilon_{r1}$& $\epsilon_{r2}$ & $\epsilon_r$ & t (mm)\\ 
\hline
 4.4 & 451 & 0.15 &0.95$\alpha$ & $\pi/4$ & 2.2 &1 & 3 &0.2\\ 
\end{tabular}
\end{ruledtabular}
\end{table}

\begin{figure}[ht!]
\centering  
\includegraphics[width=0.98\linewidth]{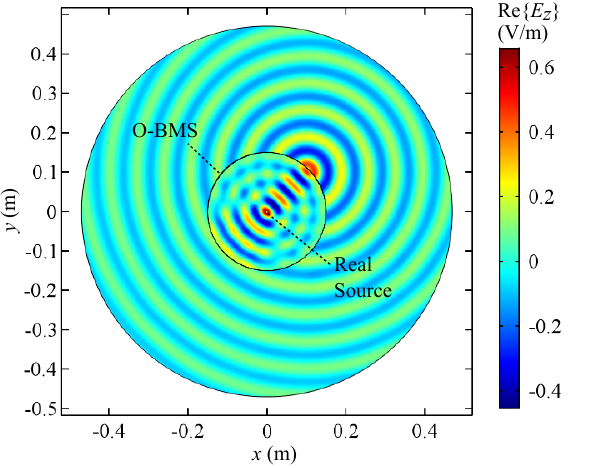}
\caption{$\mathrm{Re}\{E_z\}$ for the illusion O-BMS.}
\label{fig:ex1_fields}
\end{figure}

Next, we construct the metasurface model in COMSOL Multiphysics, in which the surface reactance layers are implemented using field-dependent surface electric current densities. The source is an electric line current placed at the origin. The simulated total (incident plus scattered) electric field distribution is depicted in Fig.~\ref{fig:ex1_fields}. Despite the complex interference pattern produced by the auxiliary internal reflections, we observe clear unperturbed wavefronts outside of the metasurface cavity emanating from the desired virtual source location.

In some scenarios, one may wish to construct a virtual source outside of the O-BMS cavity ($\rho'>\alpha$). Although at first it appears that (\ref{eqn:hankel_external_source}) should be used to obtain $\hat{E}^t_z$, we note that such a field distribution cannot actually be realized with an internally excited O-BMS. This is because (\ref{eqn:hankel_external_source}), and indeed the fields of an ideal external virtual line source, exhibit converging power flow (towards the origin) for some points on the external face of the metasurface. This can be seen from the Bessel functions which constitute the summation in (\ref{eqn:hankel_external_source}), or it can be reasoned intuitively by the fact that the Poynting vector should strictly diverge from ($\rho',\phi'$). We can circumvent this issue by approximating the desired external field distribution using (\ref{eqn:hankel_internal_source}). Following the same synthesis procedure, we would obtain an O-BMS whose external fields imitate those of the desired virtual source for the region $\rho\geq\rho'$. However, in the region $\alpha<\rho<\rho'$, the fields will not accurately depict the desired virtual source fields.

\subsection{General Penetrable Metasurface Cloak}
In this example, we design a metasurface cloak which conceals a cylindrical target from an external incident wave. Conventionally, this is often achieved with active Huygens' metasurfaces (HMS) which radiate some prescribed fields intended to destructively interfere with the scattered fields from the target~\cite{michael_cloak,paris_cloak}. While effective, this type of cloak usually requires complex circuitry to properly control and can suffer from stability issues. Alternatively, TO-MTM cloaks have also been demonstrated~\cite{MTM_cloak}. They are highly robust, capable of concealing the target from any illumination type with any incident direction. However, just like other MTM-based devices, their bulkiness significantly restricts their practical applications. 

Recently, a type of penetrable metasurface cloak has been demonstrated~\cite{Caloz_cloak}. It was proposed that if the object is penetrable by electromagnetic waves, one can synthesize a metasurface enclosure which induces zero external scattered fields while permitting some internal scattering. By carefully engineering the internal fields based on the incident fields which are known \emph{a priori}, local power conservation can be satisfied. Thus, the cloak can be realized using a passive and lossless O-BMS. An important application for this class of cloaking is the reduction of electromagnetic interference for complex wireless communications systems in which the incident field distributions are known~\cite{EMC_cloak}.

Previous literature on penetrable O-BMS cloaks assume the incident field to be a perfect plane wave. In that case, the required internal fields for satisfying local power conservation can be easily inferred to be another plane wave. Here, we extend this concept in two ways. First, we allow the incident fields to take on arbitrary forms. In general, the required internal fields for satisfying local power conservation are intricate and lack analytical descriptions. With our proposed approach, we can specify the reflected fields to be zero and numerically solve the local power conservation equation to obtain the required internal (auxiliary) fields. This straightforward procedure enables us to hide the object from more complex illuminations.

Furthermore, with some slight modifications, we can realize perfect cloaking for an impenetrable object such as a PEC cylinder. Although passive and lossless metasurfaces with such capability have been demonstrated~\cite{lossless_cloak}, they rely on carefully optimized orthogonally polarized surface waves facilitated by tensorial surface properties to achieve point-wise power balance. In contrast, our proposed approach uses a scalar metasurface which does not rely on any polarization conversion. In that light, our approach is similar to a previous work that leverages auxiliary transmitted waves to facilitate satisfaction of local power conservation in a reflective metasurface beam splitter~\cite{auxiliary}. Scalar metasurfaces can be easier to fabricate in practice compared to tensorial surfaces, owing to the simpler geometries of their constituent meta-atoms.

\begin{figure}[b!]
\centering
\includegraphics[width=0.4\textwidth]{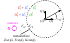}
\caption{Geometric configuration of the externally excited cylindrical O-BMS}
\label{fig:geometry_external}
\end{figure}

To develop the penetrable cloak, let us first consider the case of a general externally excited cylindrical O-BMS enclosing a homogeneous dielectric cylinder, as depicted in Fig.~\ref{fig:geometry_external}. Unlike the problem in Sec.~\ref{sec:EM_Illusion}, the incident and reflected fields now exist in the external region whereas the transmitted fields exist in the internal region, meaning the definitions in (\ref{eqn:field_vectors}) need to be adjusted accordingly. Furthermore, the BSTC equations for this problem read:
\begin{equation}
\label{eqn:external_GSTC}
\begin{split}
\frac{1}{2}(\bar{E}^i_{z}+\bar{E}^r_{z}+\bar{E}^t_{z})= &\bar{Z}_{se}\odot(\bar{H}^i_{\phi}+\bar{H}^r_{\phi}-\bar{H}^t_{\phi})\\
															  &-\bar{K}_{em}\odot(\bar{E}^i_{z}+\bar{E}^r_{z}-\bar{E}^t_{z}),\\
\frac{1}{2}(\bar{H}^i_{\phi}+\bar{H}^r_{\phi}+\bar{H}^t_{\phi}) = &\bar{Y}_{sm}\odot(\bar{E}^i_{z}+\bar{E}^r_{z}-\bar{E}^t_{z})\\
                                                                  &+\bar{K}_{em}\odot(\bar{H}^i_{\phi}+\bar{H}^r_{\phi}-\bar{H}^t_{\phi}).\\
\end{split}
\end{equation}
The new source location implies that the incident and the transmitted fields are now composed of $J_p(\cdot)$, while the reflected fields are described by $H^{(2)}_p(\cdot)$. The wavenumber for those fields also need to be changed accordingly to reflect the appropriate dielectric constants in the internal and external regions. Combining these observations, the modal admittance matrices can be inferred to be:
\begin{equation}
\label{eqn:external_impedance_matrix}
\begin{split}
\mathbf{Y}_{ex}^{\{i,t,r\}} &=\mathrm{diag}\begin{bmatrix}  	Y^{\{i,t,r\}}_{ex,p_-}(\alpha)  &
																					\cdots &
																					Y^{\{i,t,r\}}_{ex,p_+}(\alpha) \\
																					\end{bmatrix}^T,\\
Y^i_{ex,p}(\rho) &= -\frac{j\sqrt{\epsilon_{r2}}}{\eta_o}\frac{J_{p}'(k_2\rho)}{J_{p}(k_2\rho)},\\
Y^r_{ex,p}(\rho) &= -\frac{j\sqrt{\epsilon_{r2}}}{\eta_o}\frac{H^{(2)'}_{p}(k_2\rho)}{H^{(2)}_{p}(k_2\rho)},\\
Y^t_{ex,p}(\rho) &= -\frac{j\sqrt{\epsilon_{r1}}}{\eta_o}\frac{J_{p}'(k_1\rho)}{J_{p}(k_1\rho)}.\\
\end{split}																						
\end{equation}
The subscript ``$ex$" indicates that these matrices are valid for externally excited O-BMSs.

Following the same convolution-based derivation presented in Sec.~\ref{sec:analysis}, we can obtain modal transmission matrix $\hat{\mathbf{T}}_{ex}$ and modal reflection matrix $\hat{\mathbf{R}}_{ex}$ for this new configuration. Although these matrices are not explicitly used in this paper, we still present their full solutions in Appendix A for completeness; they are given by (\ref{eqn:T_matrix_ext}) and (\ref{eqn:R_matrix_ext}) respectively.

To synthesize an externally excited cylindrical O-BMS, we would again need to solve the appropriate local power conservation equation (\ref{eqn:modified_LPC}) or (\ref{eqn:modified_LPC_2}) for the required auxiliary fields. However, now we need to pay attention to substitute $\mathbf{Y}^t_{ex}$ in place of $\mathbf{Y}^t_{in}$, and $\mathbf{Y}^r_{ex}$ in place of $\mathbf{Y}^r_{in}$. After the fields everywhere are known, we use (\ref{eqn:Spectral_BSTC2_solution}) to evaluate $\{\bar{Z}_{se},\bar{Y}_{sm},\bar{K}_{em}\}$. Due to the transposed locations of the incident, transmitted and reflected fields, new field discontinuity circulant matrices $\mathbf{\Delta}_{\hat{E},ex}$ and $\mathbf{\Delta}_{\hat{H},ex}$ need to be used in place of $\mathbf{\Delta}_{\hat{E},in}$ and $\mathbf{\Delta}_{\hat{H},in}$, where
\begin{equation}
\mathbf{\Delta}_{\hat{E},ex} = -\mathbf{\Delta}_{\hat{E},in},\quad \mathbf{\Delta}_{\hat{H},ex} = -\mathbf{\Delta}_{\hat{H},in}.
\end{equation}

Last but not least, the conversion from O-BMS parameters $\{\bar{Z}_{se},\bar{Y}_{sm},\bar{K}_{em}\}$ to three-layer impedance implementation $\{\bar{Z}_i,\bar{Z}_m,\bar{Z}_o\}$ can be done with (\ref{eqn:ABCD_ZYK}) and (\ref{eqn:required_impedance_sheets}) without any modifications.

\begin{figure}[t!]
\centering  
\subfigure[]{\includegraphics[width=0.98\linewidth]{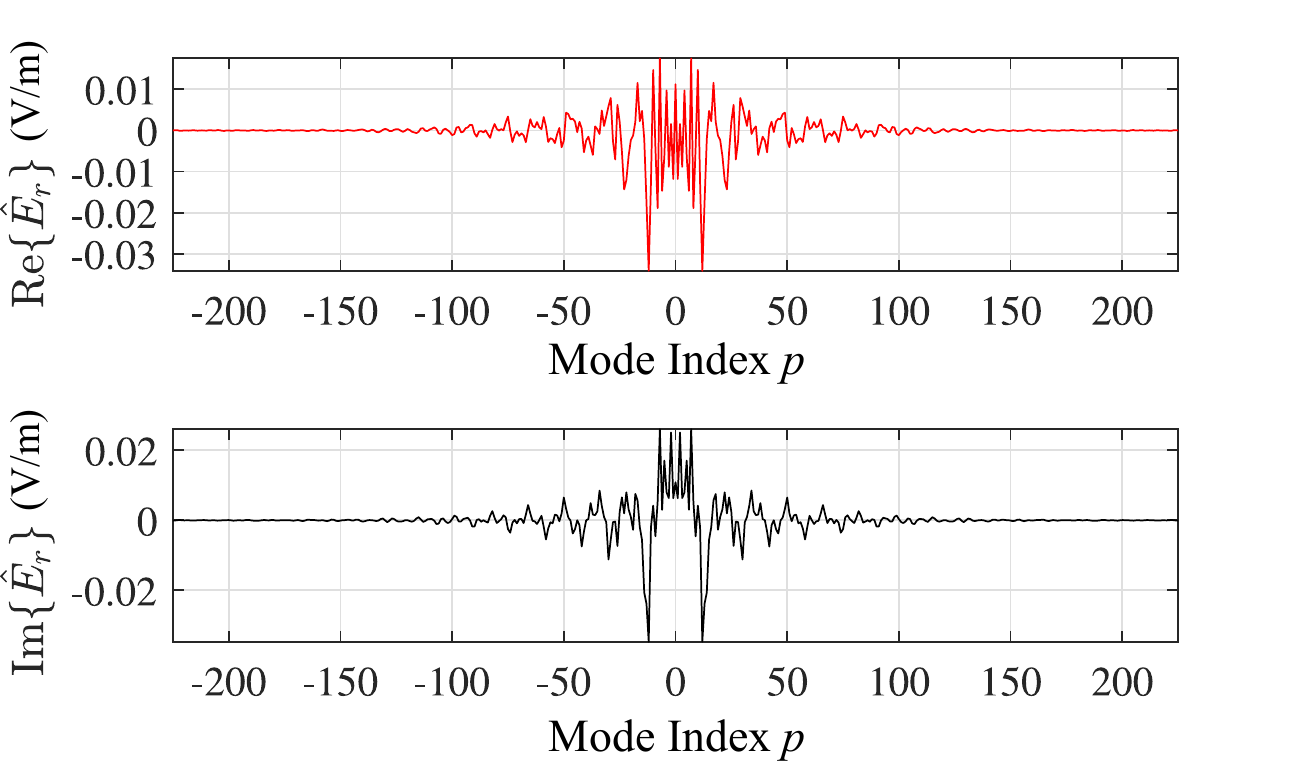}}
\subfigure[]{\includegraphics[width=0.98\linewidth]{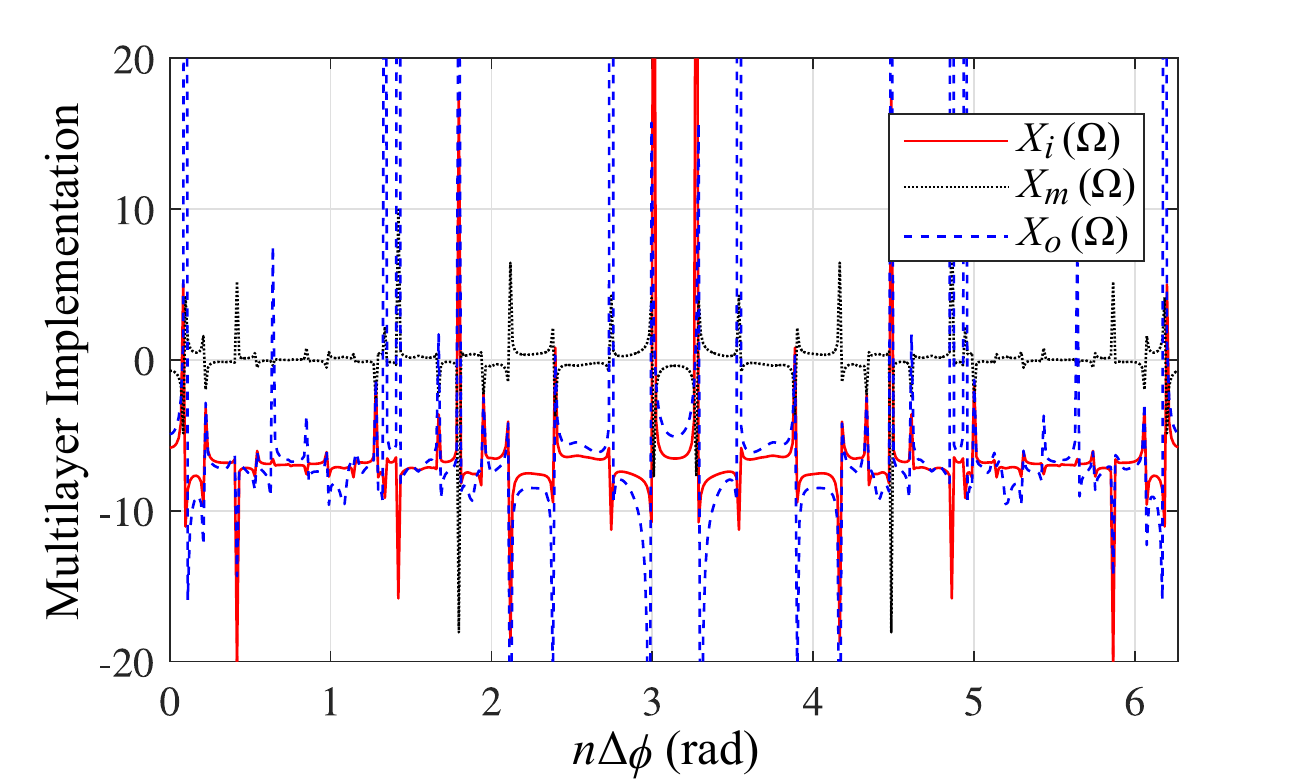}}
\caption{(a) The solved auxiliary transmitted fields for the penetrable O-BMS cloak. (b) The reactance values for the multilayer implementation of the cloak.}
\label{fig:ex3}
\end{figure}

Now, let us apply this general procedure to design a penetrable O-BMS cloak which conceals a dielectric cylinder with permittivity $\epsilon_{r1}$ from an external line source located at $(\rho,\phi)=(\rho_s,\phi_s)$. We first write the incident field modal vector with (\ref{eqn:hankel_external_source}):
\begin{equation}
\label{eqn:cloak_incident}
\hat{E}^i_z[n] = e^i_0H^{(2)}_{n^\star}(k_o\rho_s)J_{n^\star}(k_o\alpha)e^{jn^\star\phi_s},
\end{equation}
where $n^\star$ is as defined in (\ref{eqn:DFT_matrix}). The coefficient $e^i_0$ can be set to some arbitrary constant. Since the goal is to produce zero external scattered field, the desired reflected modal vector must be $\hat{E}^{r}_z=0$. As this is a reflective metasurface, we need to solve (\ref{eqn:modified_LPC_2}) for the required $\hat{E}^t_z$. Assuming the configuration as described in Table~\ref{tab:ex3_spec}, we obtain the auxiliary transmitted field modal vector to be that depicted in Fig.~\ref{fig:ex3}(a). Using these values, we obtain the three impedance layers depicted in Fig.~\ref{fig:ex3}(b). Again, the real parts are omitted since they are identically zero. 

\begin{figure}[t!]
\centering  
\subfigure[]{\includegraphics[width=0.98\linewidth]{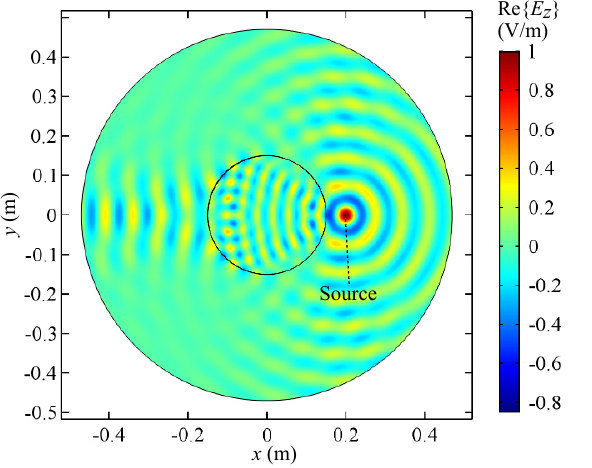}}
\subfigure[]{\includegraphics[width=0.98\linewidth]{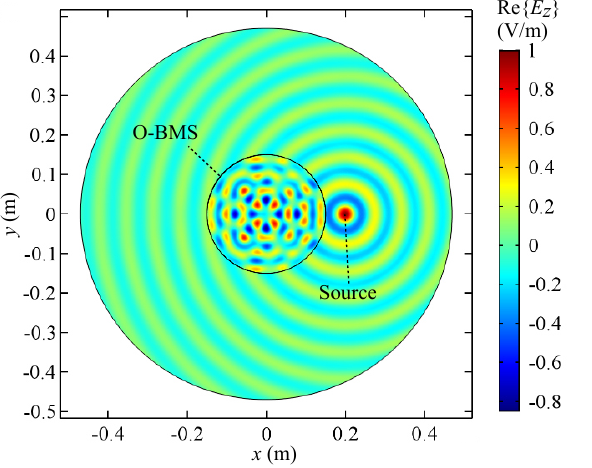}}
\caption{$\mathrm{Re}\{E_z\}$ for the dielectric cylinder (a) without the O-BMS cloak; and (b) with the O-BMS cloak.}
\label{fig:ex3_fields}
\end{figure}

To validate this design, we first simulate the scattering from the dielectric cylinder without the O-BMS cloak. The total electric field distribution is shown in Fig.~\ref{fig:ex3_fields}(a). As seen by the highly perturbed wavefronts, the cylinder interferes with the source radiation to a significant degree. Next, we add the O-BMS around the dielectric cylinder and re-simulate with the same source. The resulting fields are shown in Fig.~\ref{fig:ex3_fields}(b). Evidently, there is almost no scattering from the cylinder observable in the external region. The object would appear essentially invisible to any observer outside of the cylindrical volume.

\begin{table}[t!]
\begin{ruledtabular}
\caption{Specification for the passive lossless penetrable O-BMS cloak for concealing a dielectric cylinder}
\label{tab:ex3_spec}
\begin{tabular}{ccccccccc}
$f$ (GHz)		& $N$ 	& $\alpha$ (m)	& $\rho_s$(m) 	&$\phi_s$ (rad) & $\epsilon_{r1}$ & $\epsilon_{r2}$&$\epsilon_r$&t (mm) 	\\ 
\hline
 4.4				& 451 	& 0.15				&0.2 				& 0 					& 3 							&1&3             &0.2			\\ 
\end{tabular}
\end{ruledtabular}
\end{table}

\begin{figure}[t!]
\centering
\includegraphics[width=0.4\textwidth]{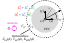}
\caption{Schematic for the penetrable O-BMS cloak for a PEC cylinder.}
\label{fig:PEC_cloak_geometry}
\end{figure}

Next, we continue the development of more advanced penetrable metasurface cloaks by designing a passive and lossless O-BMS which perfectly conceals a PEC cylinder from a known source. This can be challenging since the target itself does not permit any internal fields, meaning it is generally impossible to satisfy (\ref{eqn:modified_LPC_2}). However, if we insert a small gap between the PEC cylinder and the O-BMS, as depicted in Fig.~\ref{fig:PEC_cloak_geometry}, then both sides of the cloak can support non-zero fields. Practically, this small gap can represent an air gap or a dielectric coating around the target. There is no inherent restriction on its thickness.

We first identify the modal admittance matrices for this problem. Since the incident and reflected fields of all externally excited O-BMSs have the same constituent wave functions, we have
\begin{equation}
\mathbf{Y}_{pec}^i = \mathbf{Y}_{ex}^i,\quad\mathbf{Y}_{pec}^r = \mathbf{Y}_{ex}^r,
\end{equation}
where the subscript ``$pec$" indicates that these expressions are valid for O-BMSs surrounding a PEC cylinder.

Due to the inclusion of the PEC, the transmitted fields are now described by a linear combination of Bessel ($J_p$) and Neumann ($Y_p$) functions which satisfy $E_z(\alpha',\phi)=0$, where $\alpha'$ is the radius of the PEC cylinder. The explicit modal expansions for the fields can be written as:

\begin{equation}
\label{eqn:PEC_cloak_transmitted_fields}
\begin{split}
E^{t}_z(\rho,\phi) = \sum\limits_{p=-\infty}^{\infty}&e^t_p\Bigg[\frac{Y_p(k_1\alpha')}{d_p(k_1\alpha',k_1\alpha)}J_p(k_1\rho)\\
&-\frac{J_p(k_1\alpha')}{d_p(k_1\alpha',k_1\alpha)}Y_p(k_1\rho)\Bigg]e^{-jp\phi},\\
H^{t}_z(\rho,\phi) = \sum\limits_{p=-\infty}^{\infty}&e^t_p\frac{j\sqrt{\epsilon_{r1}}}{\eta_o}\Bigg[\frac{Y_{p}(k_1\alpha')}{d_p(k_1\alpha',k_1\alpha)}J'_p(k_1\rho)\\
&-\frac{J_p(k_1\alpha')}{d_p(k_1\alpha',k_1\alpha)}Y_p'(k_1\rho)\Bigg]e^{-jp\phi},\\
d_p(r_1,r_2)=Y_p(r_1&)J_p(r_2)-Y_p(r_2)J_p(r_1).\\
\end{split}
\end{equation}

Recalling that the modal vectors represent the N-DFT of the sampled fields at $\rho=\alpha$, we can write
\begin{equation}
\label{eqn:PEC_cloak_transmitted_electric_vector}
\begin{split}
\hat{E}^{t}_z&= \begin{bmatrix} e^t_{p_-} & \cdots& e^t_{p_+}\end{bmatrix}^T =\mathbf{\Gamma}_a\hat{E}^t_z-\mathbf{\Gamma}_b\hat{E}^t_z,\\
\mathbf{\Gamma}_a&=\mathrm{diag}\begin{bmatrix}\frac{Y_{p_-}(k_1\alpha')J_{p_-}(k_1\alpha)}{d_{p_-}(k_1\alpha',k_1\alpha)}&\cdots&\frac{Y_{p_-}(k_1\alpha')J_{p_-}(k_1\alpha)}{d_{p_-}(k_1\alpha',k_1\alpha)}\end{bmatrix}^T,\\	
\mathbf{\Gamma}_b&=\mathrm{diag}\begin{bmatrix}\frac{Y_{p_-}(k_1\alpha)J_{p_-}(k_1\alpha')}{d_{p_-}(k_1\alpha',k_1\alpha)}&\cdots&\frac{Y_{p_-}(k_1\alpha)J_{p_-}(k_1\alpha')}{d_{p_-}(k_1\alpha',k_1\alpha)}\end{bmatrix}^T,\\
\end{split}
\end{equation}
and
\begin{equation}
\label{eqn:PEC_cloak_transmitted_magnetic_vector}
\begin{split}
\hat{H}^t_z&=\mathbf{y}^t_{pec,a}\mathbf{\Gamma}_a\hat{E}^t_z-\mathbf{y}^t_{pec,b}\mathbf{\Gamma}_b\hat{E}^t_z\triangleq\mathbf{Y}^t_{pec}\hat{E}^t_z,\\
\mathbf{y}^t_{pec,a}&=-\frac{j\sqrt{\epsilon_{r1}}}{\eta_o}\cdot\mathrm{diag}\begin{bmatrix}  \frac{J_{p_-}'(k_1\alpha)}{J_{p_-}(k_1\alpha)} &\cdots &\frac{J_{p_+}'(k_1\alpha)}{J_{p_+}(k_1\alpha)}\\
\end{bmatrix}^T,\\
\mathbf{y}^t_{pec,b}&=-\frac{j\sqrt{\epsilon_{r1}}}{\eta_o}\cdot\mathrm{diag}\begin{bmatrix}  \frac{Y_{p_-}'(k_1\alpha)}{Y_{p_-}(k_1\alpha)} & \cdots &\frac{Y_{p_+}'(k_1\alpha)}{Y_{p_+}(k_1\alpha)}\\
\end{bmatrix}^T,\\	
\end{split}
\end{equation}
where $\mathbf{Y}_{pec}^t$ is the new transmitted modal admittance matrix for this particular problem. 

Noting that the BSTCs given by (32) still apply, we can reuse the expressions (\ref{eqn:T_matrix_ext}) and (\ref{eqn:R_matrix_ext}) to construct the modal transmission matrix $\mathbf{\hat{T}}_{pec}$ and modal reflection matrix $\mathbf{\hat{R}}_{pec}$ of the present configuration, simply by substituting $\mathbf{Y}^{\{i,t,r\}}_{pec}$ in place of $\mathbf{Y}^{\{i,t,r\}}_{ex}$. 

As the first step of the synthesis procedure, we find the required auxiliary $\hat{E}^t_z$ with (\ref{eqn:modified_LPC_2}), by replacing $\mathbf{Y}^t_{in}$ with $\mathbf{Y}^t_{pec}$. Since it is still an externally excited O-BMS, we can use (\ref{eqn:Spectral_BSTC2_solution}) by replacing $\mathbf{\Delta}_{\hat{E},in}$ and $\mathbf{\Delta}_{\hat{H},in}$ with $\mathbf{\Delta}_{\hat{E},ex}$ and $\mathbf{\Delta}_{\hat{E},ex}$, as is done previously for the dielectric cylinder cloak. This gives the required O-BMS parameters $\{\bar{Z}_{se},\bar{Y}_{sm},\bar{K}_{em}\}$. In the last step, we convert these parameters to the three-layer impedance implementation $\{\bar{Z}_i,\bar{Z}_m,\bar{Z}_o\}$ using (\ref{eqn:ABCD_ZYK}) and (\ref{eqn:required_impedance_sheets}) without any modifications.

\begin{figure}[hbt!]
\centering  
\subfigure[]{\includegraphics[width=0.98\linewidth]{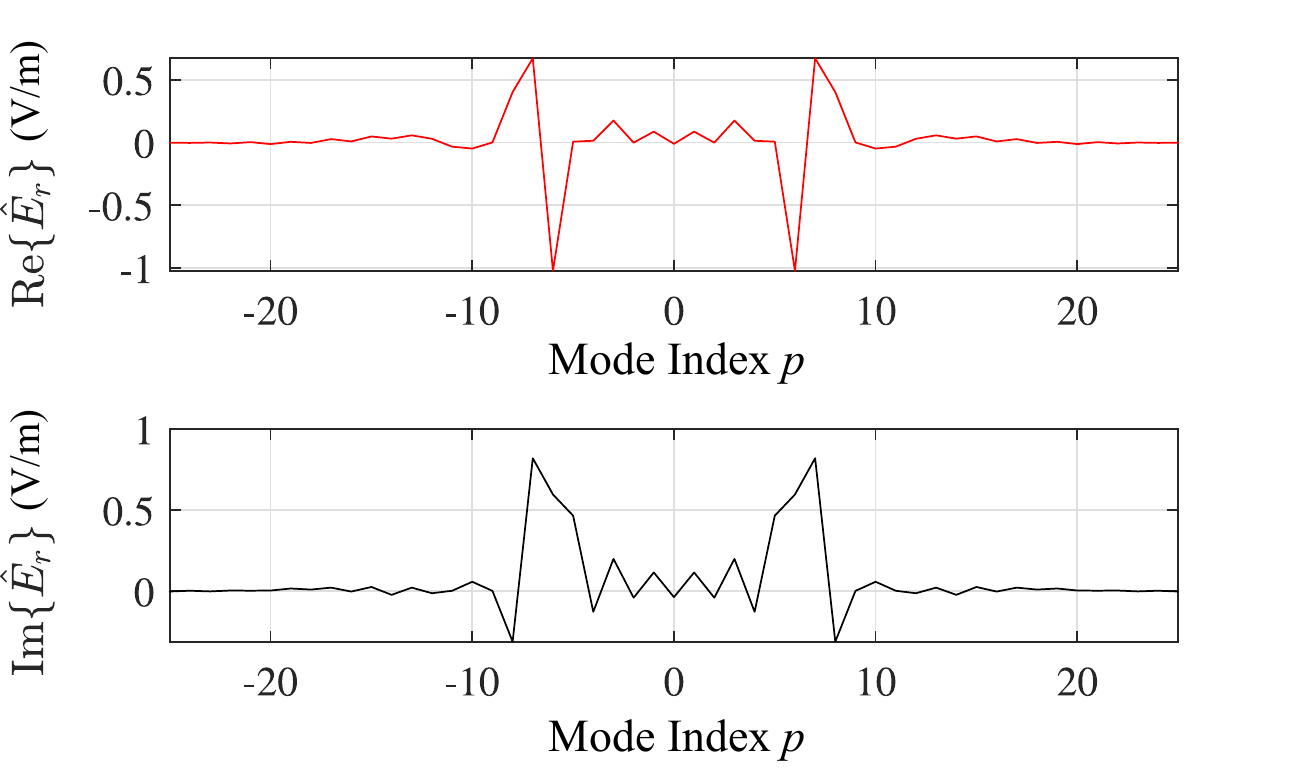}}
\subfigure[]{\includegraphics[width=0.98\linewidth]{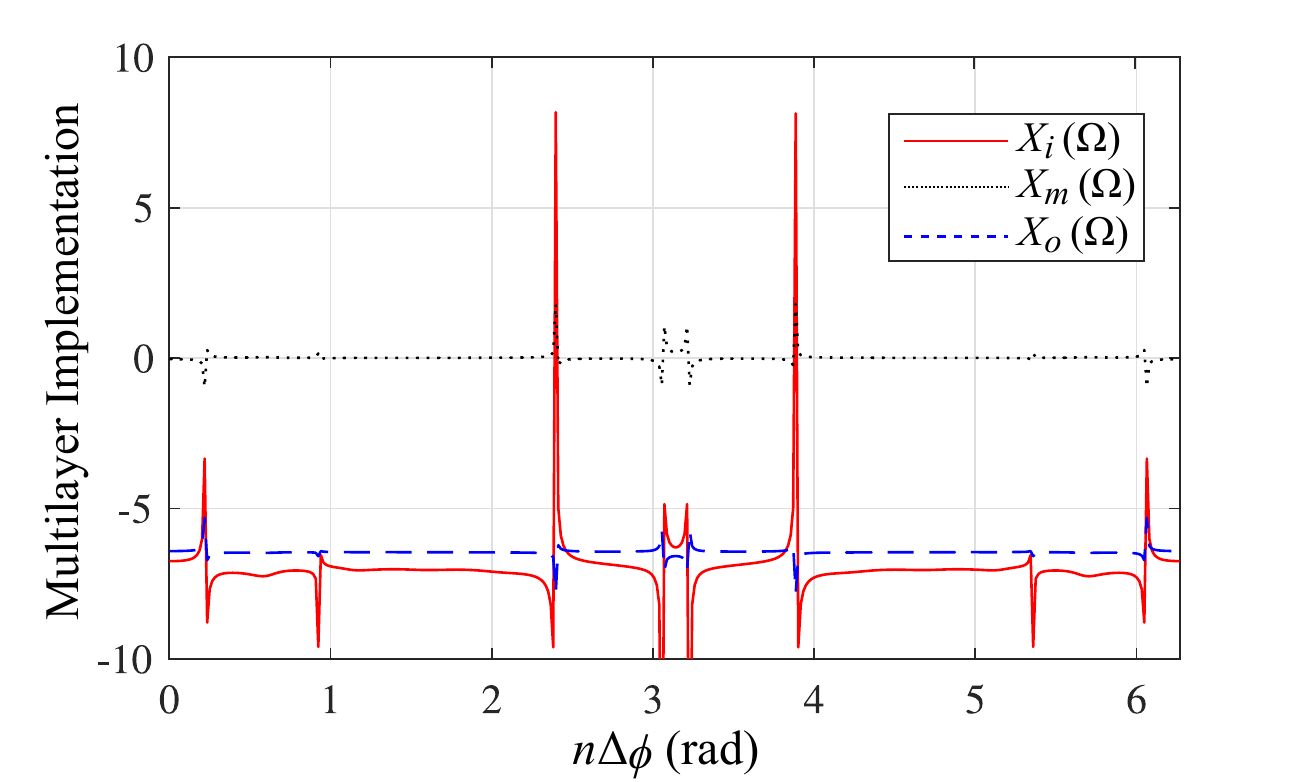}}
\caption{(a) The solved auxiliary transmitted field for the penetrable O-BMS cloak around a PEC cylinder. Higher-order modes with near-zero amplitudes are omitted. (b) The reactance values for the multilayer implementation of the cloak.}
\label{fig:ex3b}
\end{figure}

\begin{figure}[hbt!]
\centering  
\subfigure[]{\includegraphics[width=0.98\linewidth]{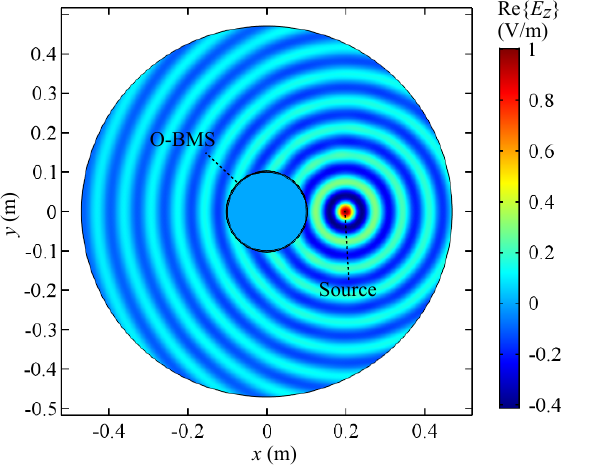}}
\subfigure[]{\includegraphics[width=0.475\linewidth]{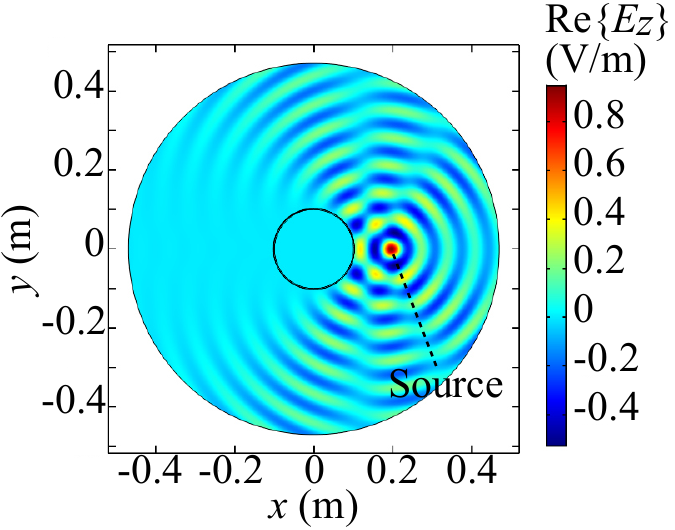}}
\subfigure[]{\includegraphics[width=0.49\linewidth]{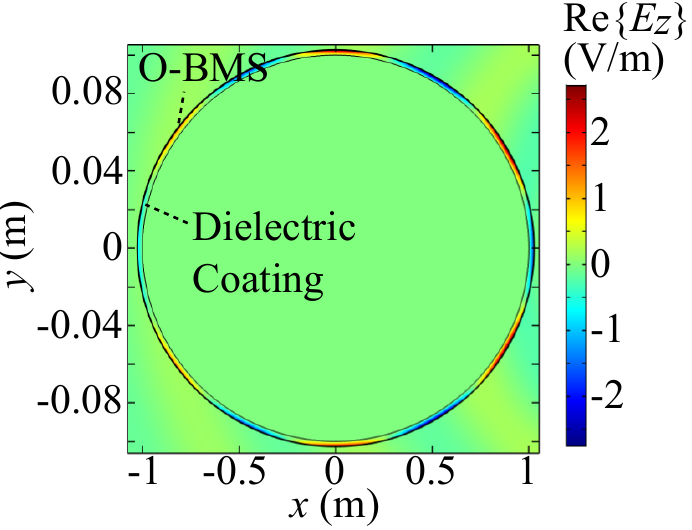}}
\caption{(a) $\mathrm{Re}\{E_z\}$ for the PEC cylinder with the O-BMS cloak. The fields inside the dielectric coating region is omitted here and plotted in (c) instead. (b) $\mathrm{Re}\{E_z\}$ for the PEC cylinder without the O-BMS cloak.}
\label{fig:ex3b_fields}
\end{figure}

Following the aforementioned procedure, we design an O-BMS cloak with the specifications outlined in Table~\ref{tab:ex3b_spec}. The solved modal vector for the required auxiliary $\hat{E}^t_z$ is shown in Fig.~\ref{fig:ex3b}(a). The corresponding passive and lossless multilayer implementation is shown in Fig.~\ref{fig:ex3b}(b). Inserting these reactance values into a COMSOL model and simulating with the designated incident fields, we obtain the electric field distribution shown in Fig.~\ref{fig:ex3b_fields}(a). Note that for better legibility, the fields inside the dielectric coating region have been excluded from this plot. The unperturbed external wavefronts signify the successful concealment of the PEC cylinder. For comparison, the fields without the O-BMS cloak is shown in Fig.~\ref{fig:ex3b}(b), in which the PEC cylinder casts a significant shadow. As illustrated in Fig.~\ref{fig:ex3b_fields}(c), the dielectric coating region around the PEC contains high-intensity standing waves corresponding to the auxiliary fields. The amplitudes of the standing waves can be reduced by increasing the coating thickness. Nevertheless, this evidently does not impair the intended functionality of the O-BMS cloak.

\begin{table}[!b]
\begin{ruledtabular}
\caption{Specification for the passive lossless O-BMS cloak for concealing a PEC cylinder}
\label{tab:ex3b_spec}
\begin{tabular}{cccccccccc}
$f$ (GHz) &$N$ & $\alpha$ (m)	&$\alpha'$ (m) &$\rho_s$ (m) 	&$\phi_s$ & $\epsilon_{r1}$& $\epsilon_{r2}$ 	&$\epsilon_r$	& t (mm)\\ 
\hline
4.4			& 401 & 0.1025&0.1&0.2 	& 0 			& 2.2 	&1						&3						&0.2\\ 
\end{tabular}
\end{ruledtabular}
\end{table}

\subsection{High-gain Cavity-excited Antenna}
In the last example, we design a cylindrical O-BMS cavity which significantly enhances the directivity of an electric line source that would otherwise produce omni-directional radiations by itself. Previously, low-profile designs with rectangular cavities have been presented~\cite{local_power_conservation,ariel_antenna}. Here, we extend this idea to a cylindrical topology which can be useful in direction-finding/navigation systems or radio beacons.

To demonstrate the use of O-BMS in constructing high-gain antennas, let us design a surface that collimates the cylindrical radiation into a directive beam towards some angle $\phi_o$. Since the incident fields are again those produced by a line source placed at the origin, the modal vector is described by (\ref{eq:ex1_inc}). The desired transmitted fields can be obtained by first identifying an envelope for its magnitude around the circumference of the cavity. For simplicity, let us assume it to be an azimuthal box function centered at $\phi=0$, with extent $\Phi$:

\begin{subnumcases}
{\label{eq:Cavity_antenna_Et_mag} \big|E_z^t(\rho=\alpha^+,\phi)\big|=}
   e_o &\quad $0<\phi\leq\frac{\Phi}{2}$, \label{eqn:Cavity_antenna_Et_mag_1}
   \\
   0 &\quad $\frac{\Phi}{2}<\phi\leq2\pi-\frac{\Phi}{2}$, \label{eqn:Cavity_antenna_Et_mag_2}
   \\
   e_o &\quad $2\pi-\frac{\Phi}{2}<\phi\leq2\pi$. \label{eqn:Cavity_antenna_Et_mag_3}
\end{subnumcases}
Note that a more sophisticated envelope function than (\ref{eq:Cavity_antenna_Et_mag}) might lead to higher maximum directivity. This is reserved for future studies.

Next, we need to properly phase the output field in order to produce the desired directional radiation. This can be done by sampling the phase of a plane wave travelling towards $\phi_o$:

\begin{equation}
\label{eq:Cavity_antenna_Et_arg} 
\angle E^t_z(\rho=\alpha^+,\phi)=-k_o\alpha\cos\big[\phi-\phi_o\big].
\end{equation}

\begin{figure}[h]
\centering  
\subfigure[]{\includegraphics[width=0.98\linewidth]{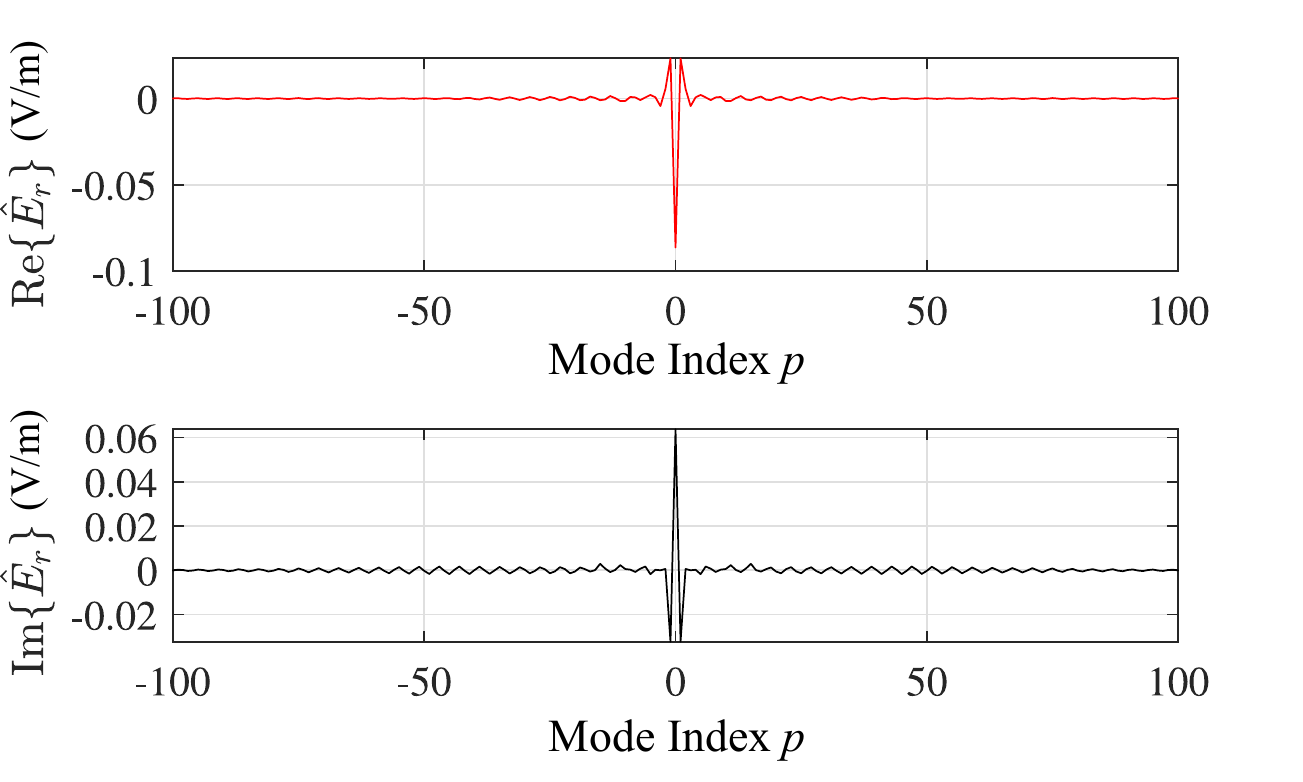}}
\subfigure[]{\includegraphics[width=0.98\linewidth]{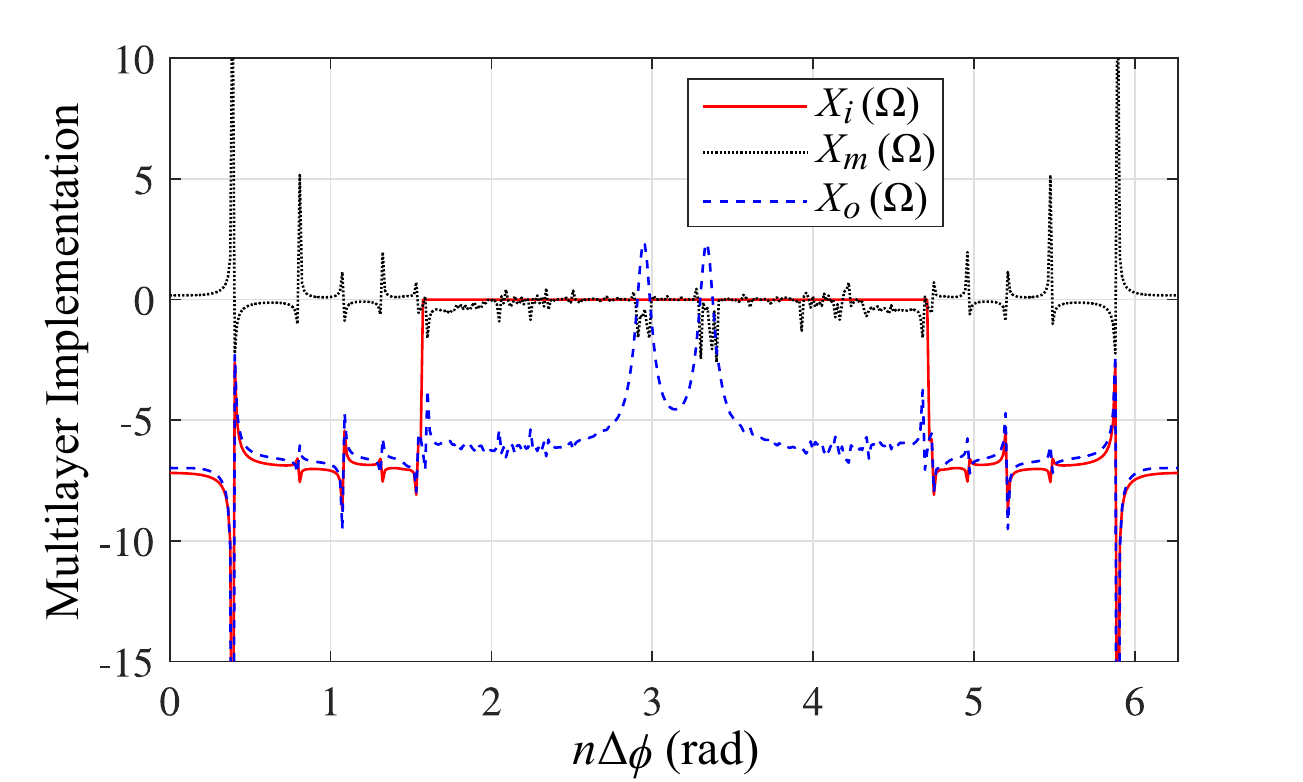}}
\caption{(a) The solved auxiliary reflected fields for the cavity-excited O-BMS antenna. Higher-order modes with near-zero amplitudes are omitted. (b) The reactance values for the multilayer implementation of the O-BMS antenna.}
\label{fig:ex4}
\end{figure}

Equations (\ref{eq:Cavity_antenna_Et_mag}) and (\ref{eq:Cavity_antenna_Et_arg}) can be used in conjunction to obtain the desired transmitted modal vector $\hat{E}^t_z$. Since this is classified as an internally excited transmissive metasurface, we solve (\ref{eqn:modified_LPC}) to find the required auxiliary reflections. Assuming the specifications are as shown in Table~\ref{tab:ex4_spec}, we calculate the reflected fields to be those depicted in Fig.~\ref{fig:ex4}(a). Correspondingly, the reactance values of the three-layer implementation of the O-BMS antenna are shown in Fig.~\ref{fig:ex4}(b). Interestingly, for the angular range $\pi/2<\phi<3\pi/2$, the inner layer has near-zero reactance, meaning that essentially half of the O-BMS cavity behaves as a PEC backing. It is consistent with intuition, since the amplitude envelope for the transmitted fields in that region is set to zero. The outer and middle layer reactance values for that region in fact have no influence on the radiation pattern of the antenna.

\begin{table}[hb!]
\begin{ruledtabular}
\caption{Specification for the cavity-excited O-BMS antenna}
\label{tab:ex4_spec}
\begin{tabular}{ccccccccc}
  $f$ (GHz) & $N$ & $\alpha$ (m) & $\phi_o $ & $\Phi$ (rad)& $\epsilon_{r1}$ &$\epsilon_{r2}$& $\epsilon_r$ & t (mm)\\ 
\hline
 4.4 & 451 & 0.15  & 0&$\pi$&1 &1& 3 &0.2\\ 
\end{tabular}
\end{ruledtabular}
\end{table}

\begin{figure}[h]
\centering  
{\includegraphics[width=0.98\linewidth]{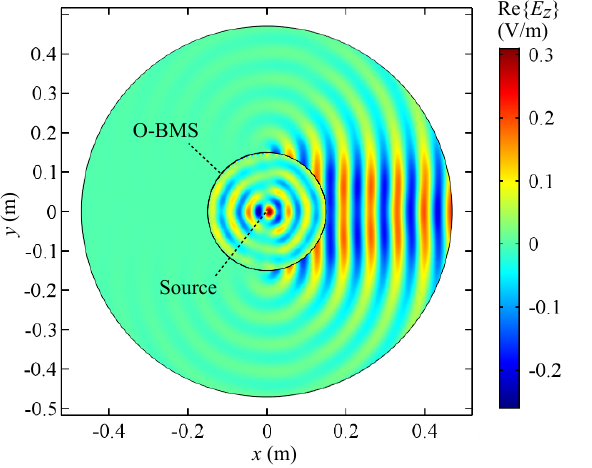}}
\caption{$\mathrm{Re}\{E_z\}$ for the cavity-excited O-BMS antenna.}
\label{fig:ex4_fields}
\end{figure}

Next, we numerically validate this design in COMSOL. In Fig.~\ref{fig:ex4_fields}, we show the simulated total electric field distribution of the antenna when it is fed by an electric line source located at the origin. The omnidirectional source field is collimated into a highly directive beam pointed at $\phi_o$.

\begin{figure}[h]
\centering  
\includegraphics[width=0.75\linewidth]{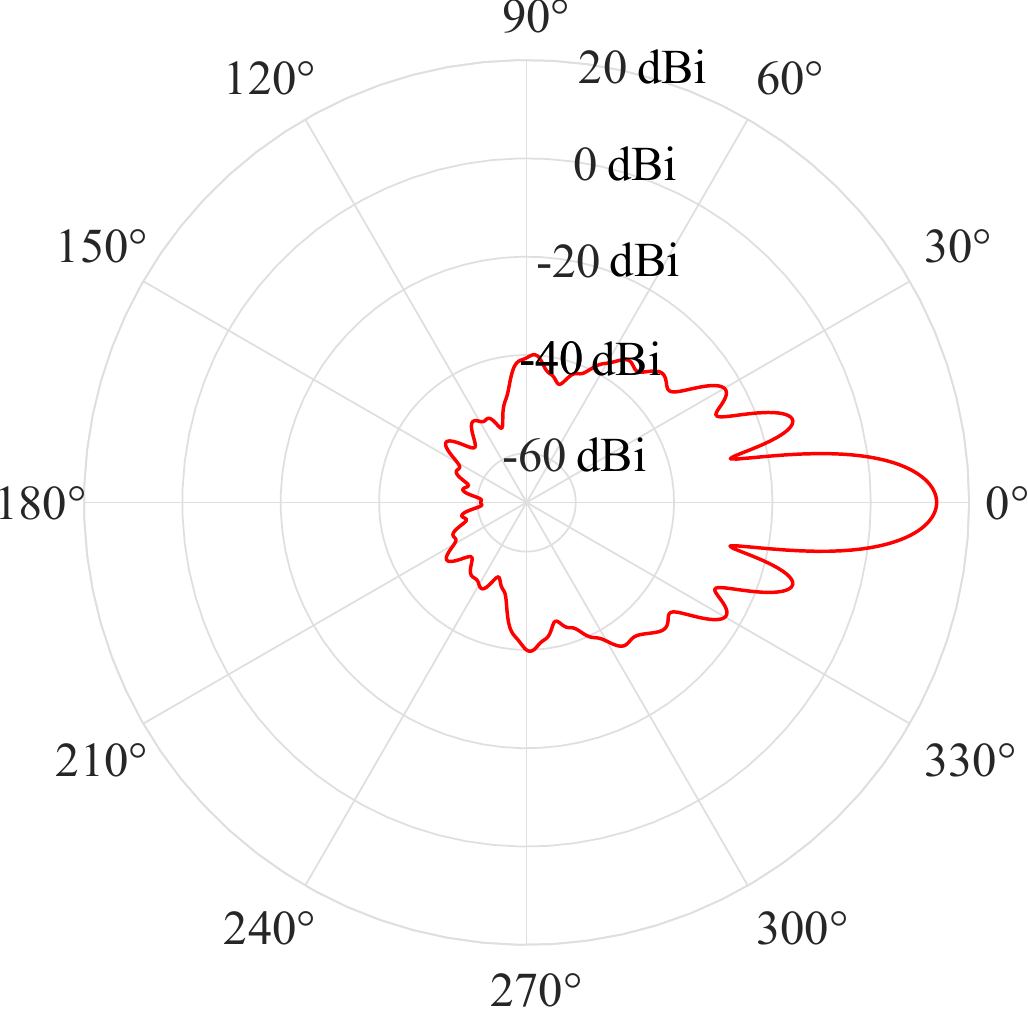}
\caption{2D directivity plot of the cavity-excited O-BMS antenna.}
\label{fig:ex4_dir}
\end{figure}

The directivity plot of the complete antenna system is shown in Fig.~\ref{fig:ex4_dir}. The maximum 2D directivity is 13.4~dBi, a significant improvement over that of the omnidirectional line source. The half power beam width is approximately 7.5$^\circ$.

\subsection{Additional Comments}
Although we have only considered TM$^\mathrm{z}$-polarized fields in this paper, it is easy to see that the proposed framework can be readily extended to TE$^\mathrm{z}$ configurations. One would simply need to modify the modal admittance matrices in accordance with the cylindrical wave functions of the TE$^\mathrm{z}$ fields. Furthermore, it is possible to model and design tensorial bianisotropic metasurfaces, which would simply involve more equations and more unknowns.

\section{Practical Realization}
\label{sec:practical_realization}
\begin{figure}[h]
\includegraphics[width=0.98\linewidth]{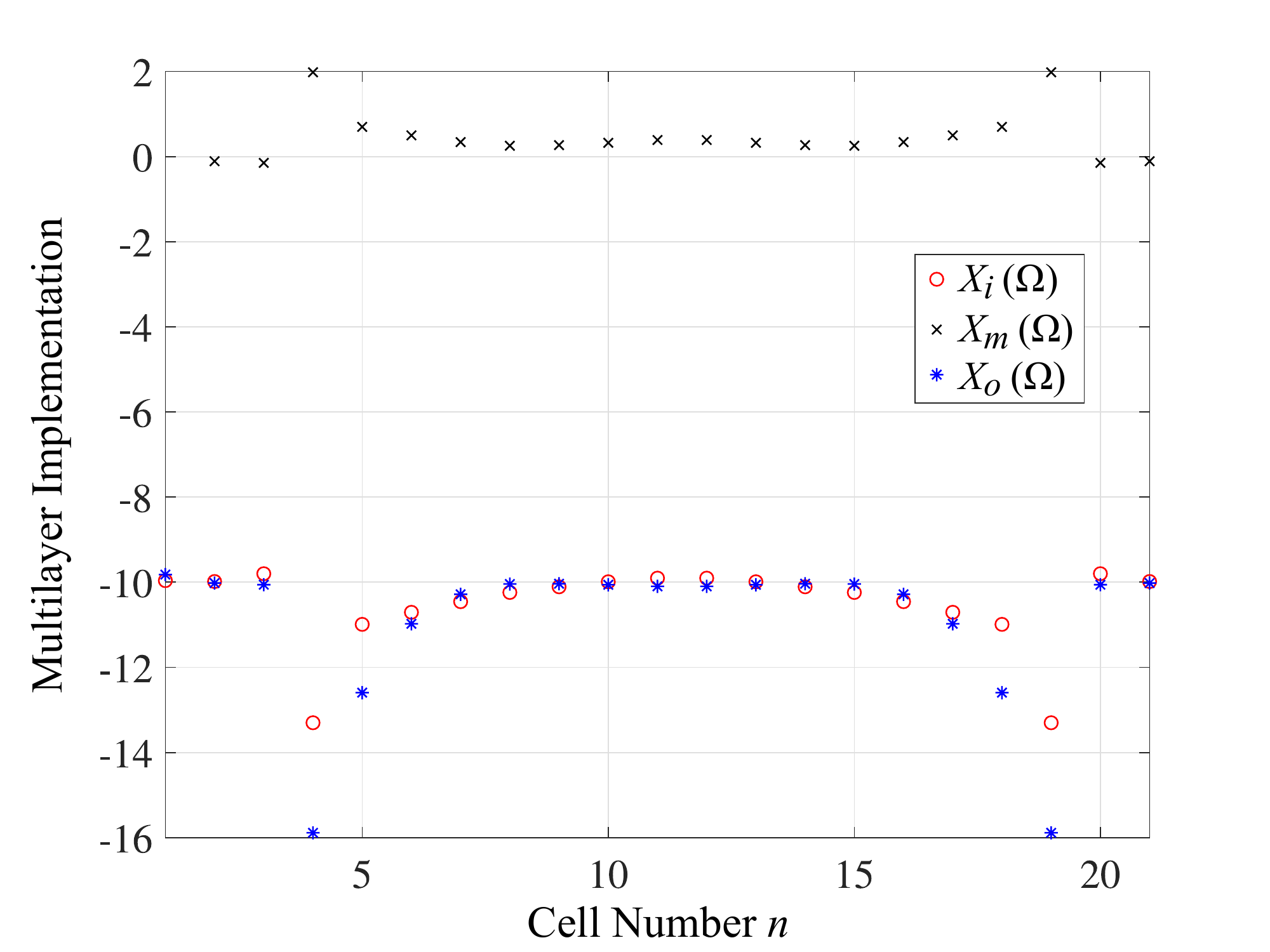}
\caption{The reactance values for the practical multilayer implementation of the illusion O-BMS.}
\label{fig:practical_design_impedances}
\end{figure}

To demonstrate the viability of the presented O-BMS designs, we discuss a possible practical implementation of the electromagnetic illusion surfaces introduced in Sec.~\ref{sec:EM_Illusion}. As remarked previously, the passive and lossless nature of our designs enables us to fabricate their constituent unit cells using three stacked layers of metallic patterns etched on two bonded sheets of PCBs. To engineer these patterns, we first assume the design specifications listed in Tab.~\ref{tab:practical_spec}, and evaluate the required reactance profiles on each layer. The resulting reactances for the $N=21$ unit cells are plotted in Fig.~\ref{fig:practical_design_impedances}. The symmetry about cell number 1 implies that only 11 unique cell designs are needed.

\begin{table}[h]
\begin{ruledtabular}
\caption{Design parameters for the practical PCB-based illusion O-BMS}
\label{tab:practical_spec}
\begin{tabular}{ccccccccc}
  $f$ (GHz) & $N$ & $\alpha$ & $\rho'$ & $\phi'$ & $\epsilon_{r1}$& $\epsilon_{r2}$ & $\epsilon_r$ & t (mm)\\ 
\hline
 10 & 21 & $\lambda_o$/2 &0.95$\alpha$ & 0 & 1 &1 & 3 &0.127\\ 
\end{tabular}
\end{ruledtabular}
\end{table}

\begin{figure}[b]
\centering  
\includegraphics[width=0.99\linewidth]{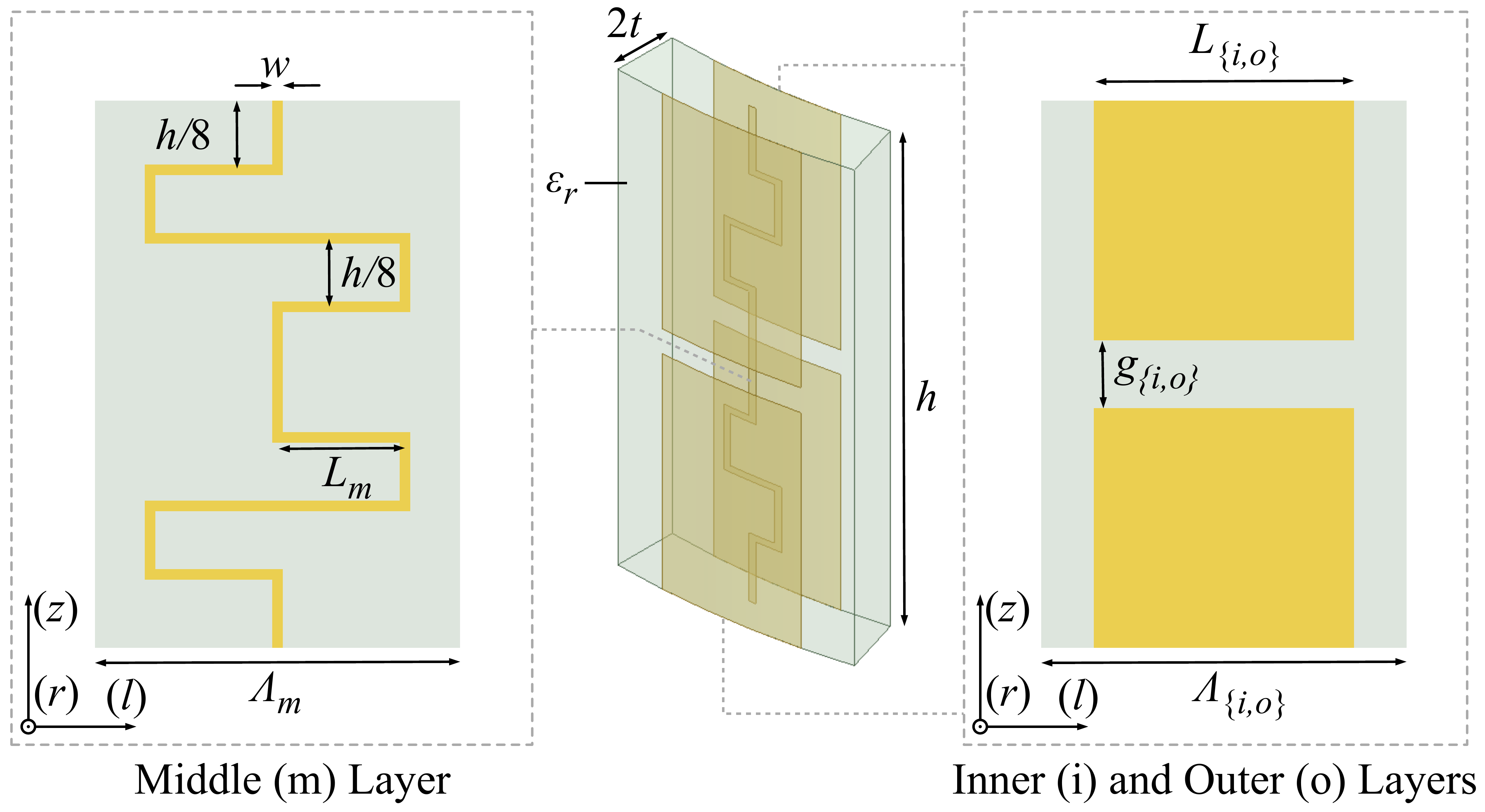}
\caption{Schematic for a type A unit cell.}
\label{fig:Cell_Type_A}
\end{figure}

\begin{figure}[t]
\centering  
\includegraphics[width=0.99\linewidth]{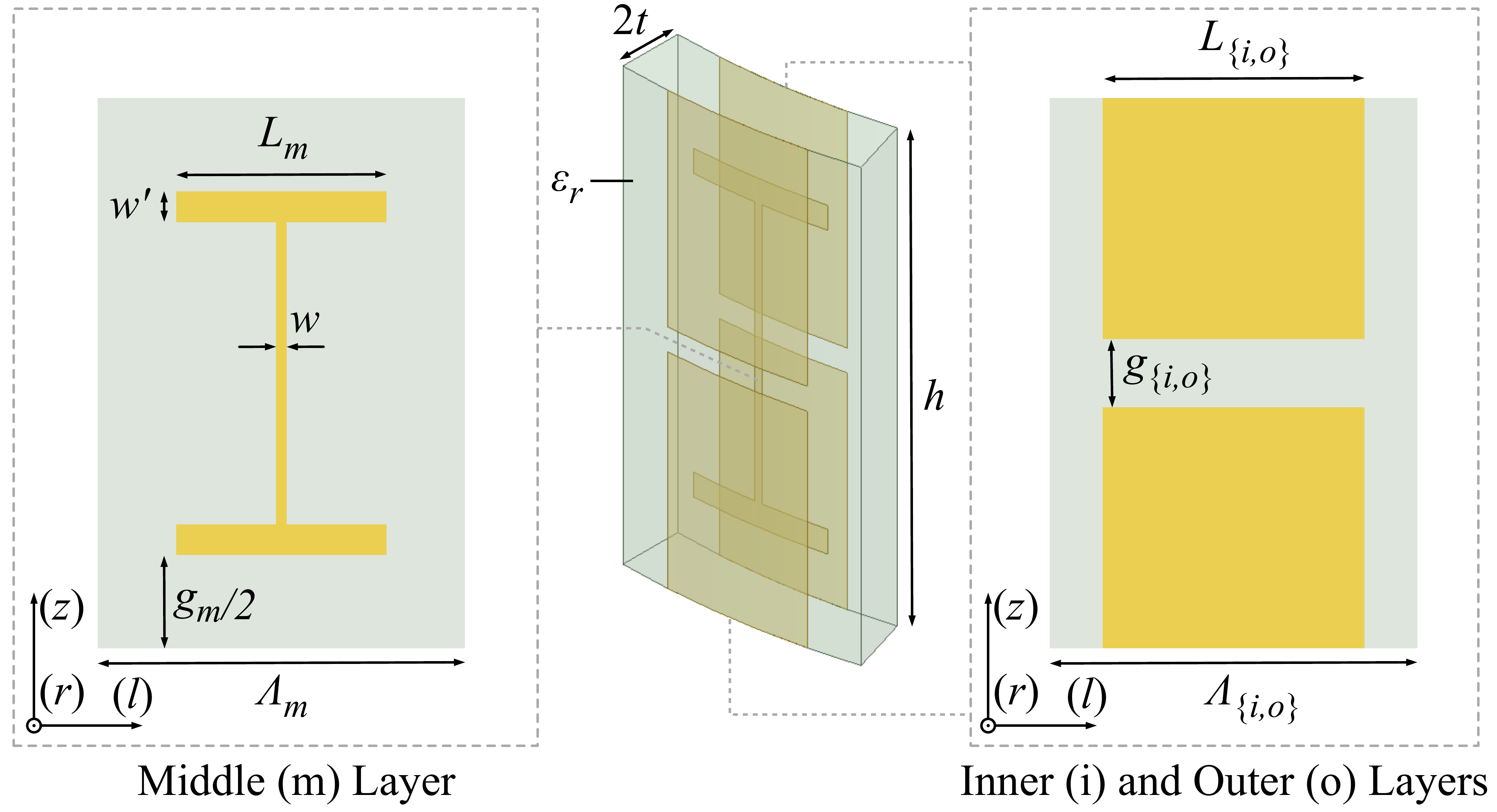}
\caption{Schematic for a type B unit cell.}
\label{fig:Cell_Type_B}
\end{figure}

We observe in Fig.~\ref{fig:practical_design_impedances} that the inner and the outer layers contain only negative reactance values, which suggests that they can be realized using capacitive gaps perpendicular to the electric field direction ($z$-axis). The middle layer can have positive or negative reactance values and thus demand more careful treatment. For unit cells with inductive $X_m$ values,  we place a meandering inductor on the middle layer, as depicted in Fig.~\ref{fig:Cell_Type_A}. This type of unit cell design will be referred to as ``type A''. For cells requiring more capacitive values of $X_m$, a loaded dipole such as that shown in Fig.~\ref{fig:Cell_Type_B} is used. These cells are referred to as ``type B''. The dipoles are used for type B cells instead of simple capacitive gaps because they bear closer resemblance to the inductors of type A cells. This may lead to better performance for the overall device due to less severe violation of the local periodic assumption. Additionally, the dipole resonances also allow type B cells to realize near-zero positive values of $X_m$ which are difficult to realize using type A cells. In both of Fig.~\ref{fig:Cell_Type_A} and Fig.~\ref{fig:Cell_Type_B}, the coordinate $l$ indicates the horizontal arc length across the cylindrical surface.

\begin{figure}[b]
\centering  
\includegraphics[width=0.95\linewidth]{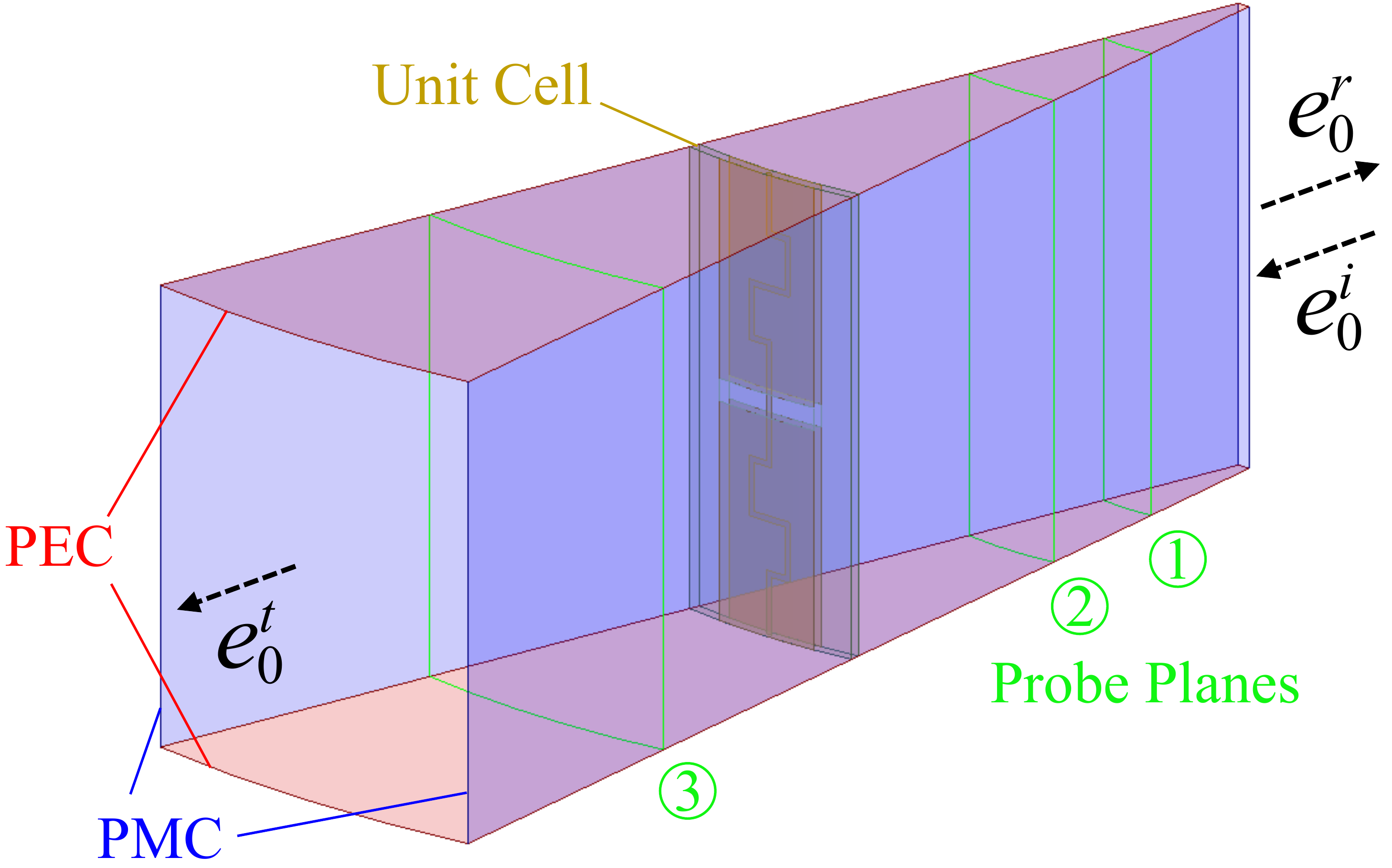}
\caption{Simulation setup used to characterize the effective $\{X_i,X_m,X_o\}$ of a three-layer O-BMS unit cell.}
\label{fig:probe_WG}
\end{figure}

To design each of the unit cells, it is necessary to establish a mapping between their geometric features and their effective reactance values. To do this, we use the simulation setup shown in Fig.~\ref{fig:probe_WG}, which was previously leveraged to design acoustic metasurfaces~\cite{angular_momentum}. In accordance with the local periodic assumption, we place a single unit cell in a wedge of a radial waveguide, and illuminate it with the $0^{th}$ mode TM${}^\mathrm{z}$ cylindrical wave. If the unit cell area is electrically small, the scattered fields, observed away from the O-BMS, will contain only the $0^{th}$ TM${}^\mathrm{z}$ mode. Thus, by evaluating the total electric fields at each of the three probe planes indicated in Fig.~\ref{fig:probe_WG}, we can obtain the coefficients $e^i_0$, $e^t_0$ and $e^r_0$. They can be used to specify the boundary electromagnetic fields in the GSTCs equations, leading to solutions for the effective \{$Z_{se}$, $Y_{sm}$, $K_{em}$\} of the unit cell under test. Lastly, we can use (\ref{eqn:required_impedance_sheets}) to extract its effective \{$X_{i}$, $X_{m}$, $X_{o}$\}.

\begin{figure}[t!]
\centering  
\includegraphics[width=0.75\linewidth]{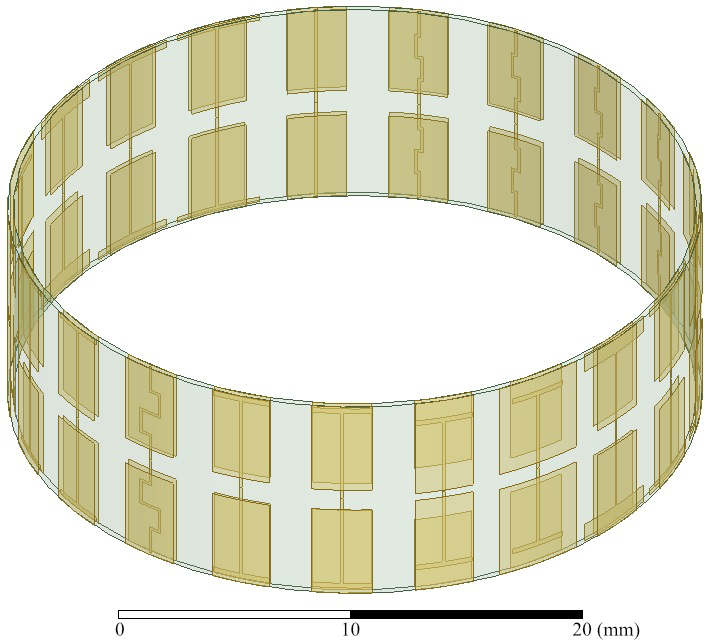}
\caption{A horizontal slice of the practical electromagnetic illusion O-BMS. The complete design consists periodic repetition of the ring along the axial direction.}
\label{fig:HFSS_Drawing}
\end{figure}

Having established a method to obtain the effective parameters of a unit cell design, we can iteratively tune its geometric features until the desired characteristics as specified in Fig~\ref{fig:practical_design_impedances} are observed. Repeating this process for the 11 unique cell designs, we arrive at the overall O-BMS illustrated in Fig.~\ref{fig:HFSS_Drawing}. Here, a single slice of the $z$-periodic device is shown. The subwavelength height of the unit cells ($h=10$mm) means that the complete design, consisting of stacked periodic repetition of the depicted slice, can be considered as homogeneous along the $z$-direction. A complete summary of the geometric parameters of this design is given in Appendix~\ref{sec:practical_geometries}.

In order to account for the impact of conductor losses which have been neglected thus far, we realize the conductive portions of the meta-atoms using metallic patterns with finite conductivity $5.8\times10^7$~S/m (copper).  We also introduce dielectric losses by setting the loss tangent of the substrate to 0.0013, which corresponds to that of the Rogers RO3003 laminate.

\begin{figure}[b]
\includegraphics[width=0.98\linewidth]{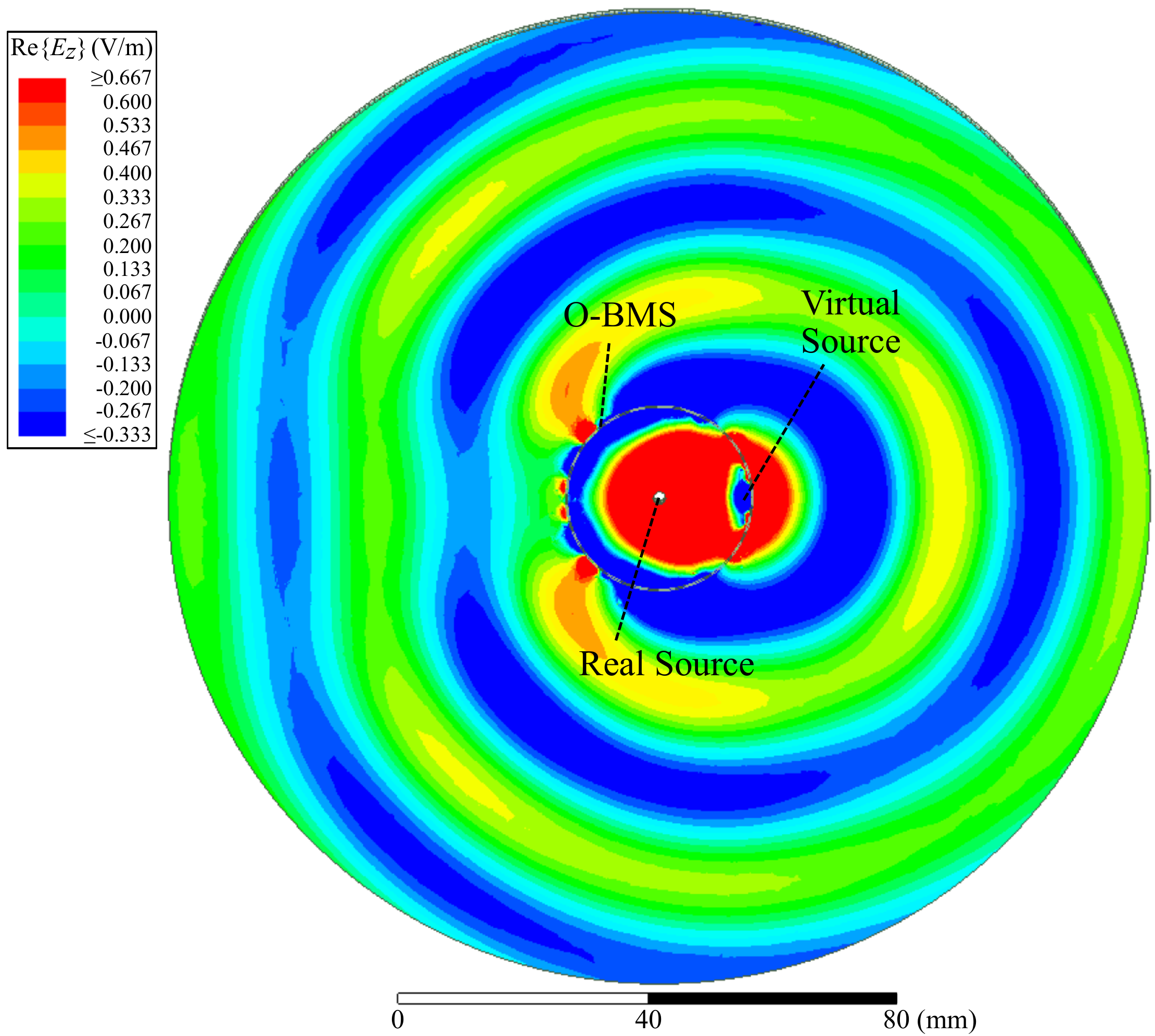}
\caption{$\mathrm{Re}\{E_z\}$ for the practical illusion O-BMS design of Fig.~\ref{fig:HFSS_Drawing}.}
\label{fig:HFSS_fields}
\end{figure}

To characterize the performance of this design, we place the slice depicted in Fig.~\ref{fig:HFSS_Drawing} into an infinitely large radial waveguide with height $h$ and excite a $z$-polarized cylindrical wave emanating from the origin. An $xy$-plane cut of the resultant total electric field distribution, obtained from full wave simulations using Ansys HFSS, is plotted in Fig.~\ref{fig:HFSS_fields}. The external fields indeed appear as if they are radiated by a virtual source located at $\pvec{\rho}'$. The slight perturbations to the otherwise cylindrical wavefronts is a result of the losses, which degrade the performance of the device by modifying its internal auxiliary reflected fields. For comparison, the fields produced by a the lossless implementation of the same design are plotted in Fig.~\ref{fig:HFSS_fields_lossless} of Appendix~\ref{sec:lossless_fields}.

It is also worth noting that the presented practical design achieves the desired functionality even without using PMC baffles. This can be attributed to the extremely thin profile of the dielectric substrates~\cite{Paul_AHMS2}. 

In conclusion, the high fidelity field transformation realized using coarsely discretized unit cells with realistic lossy meta-atoms hints at the practical viability of more sophisticated cylindrical devices such as the penetrable O-BMS cloak.

\section{Conclusion}
We presented a mode-matching framework for the analysis and synthesis of scalar cylindrical O-BMSs based on the discrete Fourier transform. By decomposing the omega-bianisotropic surface parameters into Fourier harmonics and the electromagnetic fields into cylindrical modes, we transformed the bianisotropic sheet transition conditions into simple algebraic equations which can be solved to either predict or engineer wave scattering from cylindrical metasurfaces. We also proposed a systematic procedure for designing passive and lossless cylindrical O-BMSs, which involves enforcement of local power conservation in a straightforward manner. Lastly, to bring these devices one step closer to practical realization, we present methods to realize the derived O-BMS surface parameters using a topology consisting of multiple concentric azimuthally varying electric impedance sheets.

We designed and investigated several passive and lossless scalar O-BMS-based devices including illusion metasurfaces, penetrable metasurface cloaks and high-gain metasurface antennas. Each device is numerically verified with finite element simulations, confirming the effectiveness of the proposed method. To demonstrate the practicality of the presented devices, we designed an electromagnetic illusion metasurface using realistic PCB-based meta-atoms. Its performance was validated using full wave simulations.

\section{Acknowledgment}
The authors wish to thank Nicolas Faria for the insightful discussions on cylindrical metasurfaces and their applications.

\vfill
\newpage
\bibliography{Cylindrical_OBMS_PRApplied_R2}

\appendix
\counterwithin{figure}{section}
\section{Analysis Equations for Externally Excited Cylindrical O-BMSs}
\label{sec:external_OBMS}
The modal transmission matrix $\hat{\mathbf{T}}_{ex}$ and modal reflection matrix $\hat{\mathbf{R}}_{ex}$ for the externally excited TM$^\mathrm{z}$-polarized O-BMS pictured in Fig.~\ref{fig:geometry_external} are as follows:
\begin{footnotesize}
\begin{equation}
\label{eqn:T_matrix_ext}
\begin{split}
\mathbf{\hat{T}}_{ex} =& \mathbf{\hat{t}}_{ex,b}^{-1}\mathbf{\hat{t}}_{ex,a},\\
\mathbf{\hat{t}}_{ex,a} =& \bigg(\frac{1}{2}\mathbf{I}+\mathbf{\hat{K}}-\mathbf{\hat{Z}}\mathbf{Y}_{ex}^r\bigg)^{-1}\bigg(\frac{1}{2}\mathbf{I}+\mathbf{\hat{K}}-\mathbf{\hat{Z}}\mathbf{Y}_{ex}^i\bigg)\\&-\bigg(\frac{1}{2}\mathbf{Y}_{ex}^r-\mathbf{\hat{Y}}-\mathbf{\hat{K}}\mathbf{Y}_{ex}^r\bigg)^{-1}\bigg(\frac{1}{2}\mathbf{Y}_{ex}^i-\mathbf{\hat{Y}}-\mathbf{\hat{K}}\mathbf{Y}_{ex}^i\bigg),\\
\mathbf{\hat{t}}_{ex,b} =& \bigg(\frac{1}{2}\mathbf{Y}_{ex}^r-\mathbf{\hat{Y}}-\mathbf{\hat{K}}\mathbf{Y}_{ex}^r\bigg)^{-1}\bigg(\frac{1}{2}\mathbf{Y}_{ex}^t+\mathbf{\hat{Y}}+\mathbf{\hat{K}}\mathbf{Y}_{ex}^t\bigg)\\
&-\bigg(\frac{1}{2}\mathbf{I}+\mathbf{\hat{K}}-\mathbf{\hat{Z}}\mathbf{Y}_{ex}^r\bigg)^{-1}\bigg(\frac{1}{2}\mathbf{I}-\mathbf{\hat{K}}+\mathbf{\hat{Z}}\mathbf{Y}_{ex}^t\bigg).
\end{split}
\end{equation}
\end{footnotesize}
\begin{footnotesize}
\begin{equation}
\label{eqn:R_matrix_ext}
\begin{split}
\mathbf{\hat{R}}_{ex} =& \mathbf{\hat{r}}_{ex,b}^{-1}\mathbf{\hat{r}}_{ex,a},\\
\mathbf{\hat{r}}_{ex,a} =& \bigg(\frac{1}{2}\mathbf{I}-\mathbf{\hat{K}}+\mathbf{\hat{Z}}\mathbf{Y}_{ex}^t\bigg)^{-1}\bigg(\frac{1}{2}\mathbf{I}+\mathbf{\hat{K}}-\mathbf{\hat{Z}}\mathbf{Y}_{ex}^i\bigg)\\
&-\bigg(\frac{1}{2}\mathbf{Y}_{ex}^t+\mathbf{\hat{Y}}+\mathbf{\hat{K}}\mathbf{Y}_{ex}^t\bigg)^{-1}\bigg(\frac{1}{2}\mathbf{Y}_{ex}^i-\mathbf{\hat{Y}}-\mathbf{\hat{K}}\mathbf{Y}_{ex}^i\bigg),\\
\mathbf{\hat{r}}_{ex,b} =& \bigg(\frac{1}{2}\mathbf{Y}_{ex}^t+\mathbf{\hat{Y}}+\mathbf{\hat{K}}\mathbf{Y}_{ex}^t\bigg)^{-1}\bigg(\frac{1}{2}\mathbf{Y}_{ex}^r-\mathbf{\hat{Y}}-\mathbf{\hat{K}}\mathbf{Y}_{ex}^r\bigg)\\
&-\bigg(\frac{1}{2}\mathbf{I}-\mathbf{\hat{K}}+\mathbf{\hat{Z}}\mathbf{Y}_{ex}^t\bigg)^{-1}\bigg(\frac{1}{2}\mathbf{I}+\mathbf{\hat{K}}-\mathbf{\hat{Z}}\mathbf{Y}_{ex}^r\bigg).
\end{split}
\end{equation}
\end{footnotesize}

To obtain the the modal transmission matrix $\mathbf{\hat{T}}_{pec}$ and modal reflection matrix $\mathbf{\hat{R}}_{pec}$, for an O-BMS surrounding a PEC cylinder, replace $\mathbf{Y}_{ex}^{\{i,t,r\}}$ with $\mathbf{Y}_{pec}^{\{i,t,r\}}$.

\section{Geometric Parameters for the Practical Illusion O-BMS Design}
\label{sec:practical_geometries}
The 11 unique unit cell geometries utilized by the design depicted in Fig.~\ref{fig:HFSS_Drawing} are summarized here. The features listed in Tab.~\ref{tab:geometries_common} are shared among all cells. 

\begin{table}[h]
\centering
\caption{Common geometric parameters (in millimeters) shared among all unit cells in Fig.~\ref{fig:HFSS_Drawing}.}
\label{tab:geometries_common}
\begin{tabular}{cccccc}
\hline\hline
   $\Lambda_i$ & $\Lambda_m$ & $\Lambda_o$ & $h$ & $w$ & $t$ \\ 
\hline
4.485 & 4.523 &  4.561 & 10.00 &  0.132 & 0.127  \\ 
\hline\hline
\end{tabular}
\end{table}

The cell-specific parameters are summarized in the following tables, where the cells are grouped into type A (Tab.~\ref{tab:geometries_type_A}) and type B (Tab.~\ref{tab:geometries_type_B}) as defined in Sec.~\ref{sec:practical_realization}. 

\begin{table}[h]
\centering
\caption{Specific geometric parameters (in millimeters) for the type A unit cells in Fig.~\ref{fig:HFSS_Drawing}.}
\label{tab:geometries_type_A}
\begin{tabular}{cccccc}
\hline\hline
   Cell \# & $L_i$ & $L_m$& $L_o$ & $g_i$ & $g_o$ \\ 
\hline
1 &  2.616 &  0.508 & 2.661 & 1.220 & 1.336  \\ 
9, 14 &  2.616 & 0.000 & 2.661 &  1.220 & 1.330  \\
10, 13 &  2.616 & 0.125 &2.661 & 1.344 & 1.270  \\  
11, 12 & 2.616 &  0.168 &2.661 &  1.375 &  1.230 \\  
\hline\hline
\end{tabular}
\end{table}

\begin{table}[h]
\centering
\caption{Specific geometric parameters (in millimeters) for the type B unit cells in Fig.~\ref{fig:HFSS_Drawing}.}
\label{tab:geometries_type_B}
\begin{tabular}{cccccccc}
\hline\hline
   Cell \# & $L_i$ & $L_m$ & $L_o$ & $g_i$ & $g_m$ & $g_o$ & $w'$ \\ 
\hline
2, 21 & 2.616 & 2.638 & 2.661 & 1.400 & 0.262 & 1.190 & 0.250\\ 
3, 20 & 2.616 & 2.638 & 2.661 & 1.430 & 0.380 & 1.070 & 0.250\\
4, 19 & 2.616 & 2.427 & 2.661 & 2.780 & 2.000 & 0.350 & 0.250\\  
5, 18 & 2.616 & 2.638 & 3.991 & 1.827 & 1.680 & 0.800 & 0.250\\  
6, 17 & 2.616 & 3.958 & 2.661 & 1.780 & 0.000 & 1.328 & 0.900\\  
7, 16 & 2.616 & 3.958 & 2.661 & 1.354 & 0.000 & 1.414 & 0.400\\ 
8, 15 & 2.616 & 3.958 & 2.661 & 1.295 & 0.000 & 1.375 & 0.250\\   
\hline\hline
\end{tabular}
\end{table}
\vfill
\newpage
\section{Simulation Results of Lossless Illusion O-BMS}
\label{sec:lossless_fields}
\begin{figure}[h]
\includegraphics[width=0.98\linewidth]{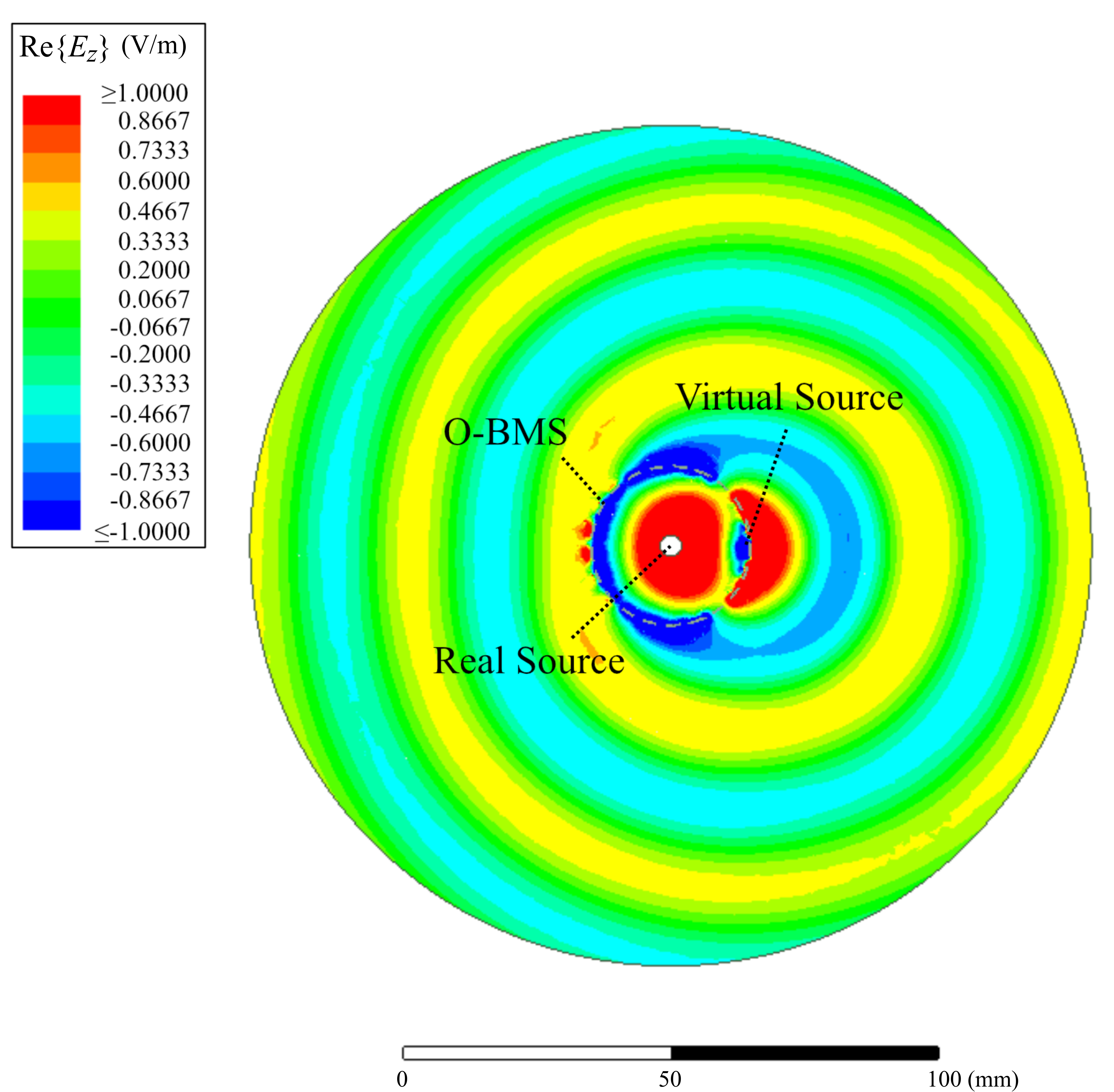}
\caption{Simulated $\mathrm{Re}\{E_z\}$ for the practical illusion O-BMS design of Fig.~\ref{fig:HFSS_Drawing}, neglecting conduction and dielectric losses.}
\label{fig:HFSS_fields_lossless}
\end{figure}

\end{document}